\documentclass[prb,twocolumn,showpacs,floatfix]{revtex4-1}
\usepackage{epsfig,color}
\usepackage{amsmath}

\begin{document}
\title{Application of the Landau-Zener-St{\"u}ckelberg-Majorana dynamics in an electrically driven flip of a hole spin}
\author{W. J. Pasek} \affiliation{University of Campinas, School of Applied Sciences, \\
R. Pedro Zaccaria, 1300 - Jd. Santa Luiza, Limeira - SP, 13484-350, Brazil}
\author{M. Z. Maialle} \affiliation{University of Campinas, School of Applied Sciences, \\
R. Pedro Zaccaria, 1300 - Jd. Santa Luiza, Limeira - SP, 13484-350, Brazil}
\author{M. H. Degani} \affiliation{University of Campinas, School of Applied Sciences, \\
R. Pedro Zaccaria, 1300 - Jd. Santa Luiza, Limeira - SP, 13484-350, Brazil}
\date{\today}

\begin{abstract}
An idea of employing the Landau-Zener-St{\"u}ckelberg-Majorana (LZSM) dynamics to flip a spin of a single ground state hole is introduced and explored by a time-dependent simulation. This~configuration interaction study considers a hole confined in a quantum molecule formed in InSb $\langle111\rangle$ quantum wire by application of an electrostatic potential. An up-down spin-mixing avoided crossing is formed by non-axial terms in the Kohn-Luttinger Hamiltonian and the Dresselhaus spin-orbit one. Manipulation of the system is possible by dynamic change of external vertical electric field, which enables the consecutive driving of the hole through two anticrossings. Moreover, a simple model of the \textit{power-law} type noise that impedes the precise electric control of the system is included in the form of random telegraph noise to estimate the limitations of the working conditions. We~show that in principle the process is possible, but it requires a precise control of parameters of the driving impulse.
\end{abstract}

\maketitle

\section{Introduction}

The electric dipole spin resonance (EDSR) is a process in which the spin state of a quantum system is manipulated by the means of an ac electric field.\cite{edsr1,edsr2,edsr3,edsr4,edsr5,edsr6,edsr7,edsr8,edsr9,edsr10} This process can utilize a few different mechanisms of electric-spin coupling: the spin-orbit interaction,\cite{edsr1,edsr3,edsr6,edsr7,edsr8,edsr10} the spatial inhomogeneity of the applied magnetic field\cite{edsr2,edsr5} or of the hyperfine interaction.\cite{edsr4} If the frequency of the electric signal is resonant to the relevant energy difference of two levels with different spin, then a transition between the levels may be induced, depending on a certain set of selection rules.

A typical EDSR transition is done between two uncoupled spin states. However, when two levels are involved in an avoided crossing, then driving the system through this anticrossing is described by the Landau-Zener dynamics instead. When the driving is periodic, the system accumulates the St{\"u}ckelberg phase between the transitions and this leads to a constructive or destructive interference, depending on the specific parameters of a given system. The theory related to systems of this kind is described in a review article of Ref. \onlinecite{LZ_theory}.

Multiple harmonic generation in EDSR in an InAs nanowire double quantum dot was recently observed for conduction band electrons in double quantum dots.\cite{LZSM_elec_experiment} The harmonics display a remarkable detuning dependence: near the interdot charge transition as many as eight harmonics were observed, while at large detunings only the fundamental spin resonance condition was detected. In following theoretical studies the transport dynamics of a periodically driven system, modeling the level structure of a two-electron double quantum dot, was studied.\cite{LZSM_elec_theory,rudner} It was shown that the observed multiphoton resonances, which are dominant near interdot charge transitions, are due to multilevel Landau-Zener-St{\"u}ckelberg-Majorana interference. The main features observed in the experiments of Ref.~\onlinecite{LZSM_elec_experiment} were replicated: multiphoton resonances up to eight photons, a robust odd-even dependence, and oscillations in the electric dipole spin-resonance signal as a function of energy-level detuning.

The Landau-Zener dynamic was used to study the possibility of manipulating the S-T+ avoided crossing that arises due to the hyperfine interactions in a system of two electrons in a double quantum dot in GaAs.\cite{burkard-1,burkard-2} The results concern a two electron Landau-Zener system with the spin-mixing singlet-triplet avoided crossing, resulting from the hyperfine interaction. In the both works, the necessity of going beyond the simplest infinite time Landau-Zener model is stressed out, and the finite-time Landau-Zener theory is employed. Moreover, the formulated master-equation formalism allowed to study the impact of phonon-mediated hyperfine relaxation and charge-noise-induced dephasing on the evolution of the system.\cite{burkard-2} In the corresponding experimental work, Ref. \onlinecite{Petta}, an all-electrical method for quantum control was presented that relies on electron-nuclear spin coupling and drives and drives spin rotations on nanosecond time scales. Interference patterns were observed\cite{Petta} in singlet-state occupation as a function of waiting time between consecutive sweeping the system back and forth through a singlet-triplet avoided crossing, due to phase accumulation, with agreement in the Landau-Zener theory.

In a recent work, a p-channel silicon metal-oxide-semiconductor field-effect transistor with a double dot in the channel, formed by a pair of defects or impurities, was studied.\cite{Ono} A two-spin EDSR was realized experimentally, with the main line as well as additional few-photon lines visible. A supression of the spin resonance was found in the viccinity of a singlet-triplet avoided crossing.

\section{Motivation}

The EDSR manipulation scheme was realized for valence band holes in a quantum molecule created in a gated InSb nanowire.\cite{holes_exp} The mentioned work employed the strong spin-orbit coupling of this material for the spin flipping and measured the transport through the system as a result of lifting the Pauli spin blockade.

\begin{figure}[ht!]
\epsfxsize=70mm \epsfbox[75 40 680 540] {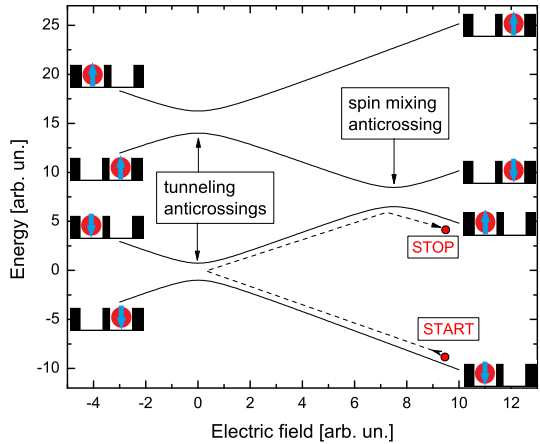}
\caption{An ilustrative scheme of the transfer proces idea. The spin-dependent energy level structure is shown with the desired spin flipping process utilizing two avoided crossings.}\label{scheme}
\end{figure}

In this work we suggest a scheme for reversing the spin state of a single hole confined in two quantum dots with tunneling coupling, whose energy level structure is presented schematically in Fig. \ref{scheme}. We have taken advantage of an avoided crossing involving the states localized in the same dot but with an opposing spin characteristics. We~propose an electrical control signal that leads to the transition from the ground state, of the heavy hole spin-down type, to the first excited state, of the heavy hole spin-up type. Instead of using a long periodic signal, as typically done in EDSR processes, the system is driven through two anticrossings only a few times by a short cosine impulse. High transition efficiency of about $0.99$ is obtained by the detailed balancing of the impulse parameters. The process is one order of magnitude faster than the alternative EDSR realized in the same system, which is especially important in the context of limited spin coherence time when performing spin operations.

\section{Theory}

\subsection{Geometry of the system}

We consider a system of a single hole in a quantum molecule consisting of two quantum dots coupled vertically. The dots are made by applying an external electrostatic potential to an InSb quantum wire of a $\langle111\rangle$ zincblende crystal structure. The nanowire is assumed to have circular shape in cross-section and a radius of \mbox{$R_{dot}=50$ nm}. The geometry of the system is similar to the one considered in Ref. \onlinecite{holes_exp}.

\begin{figure*}[ht!]
\begin{tabular}{cc}
(a) & (b)\\
\rotatebox{0}{\epsfxsize=70mm \epsfbox[65 35 705 505] {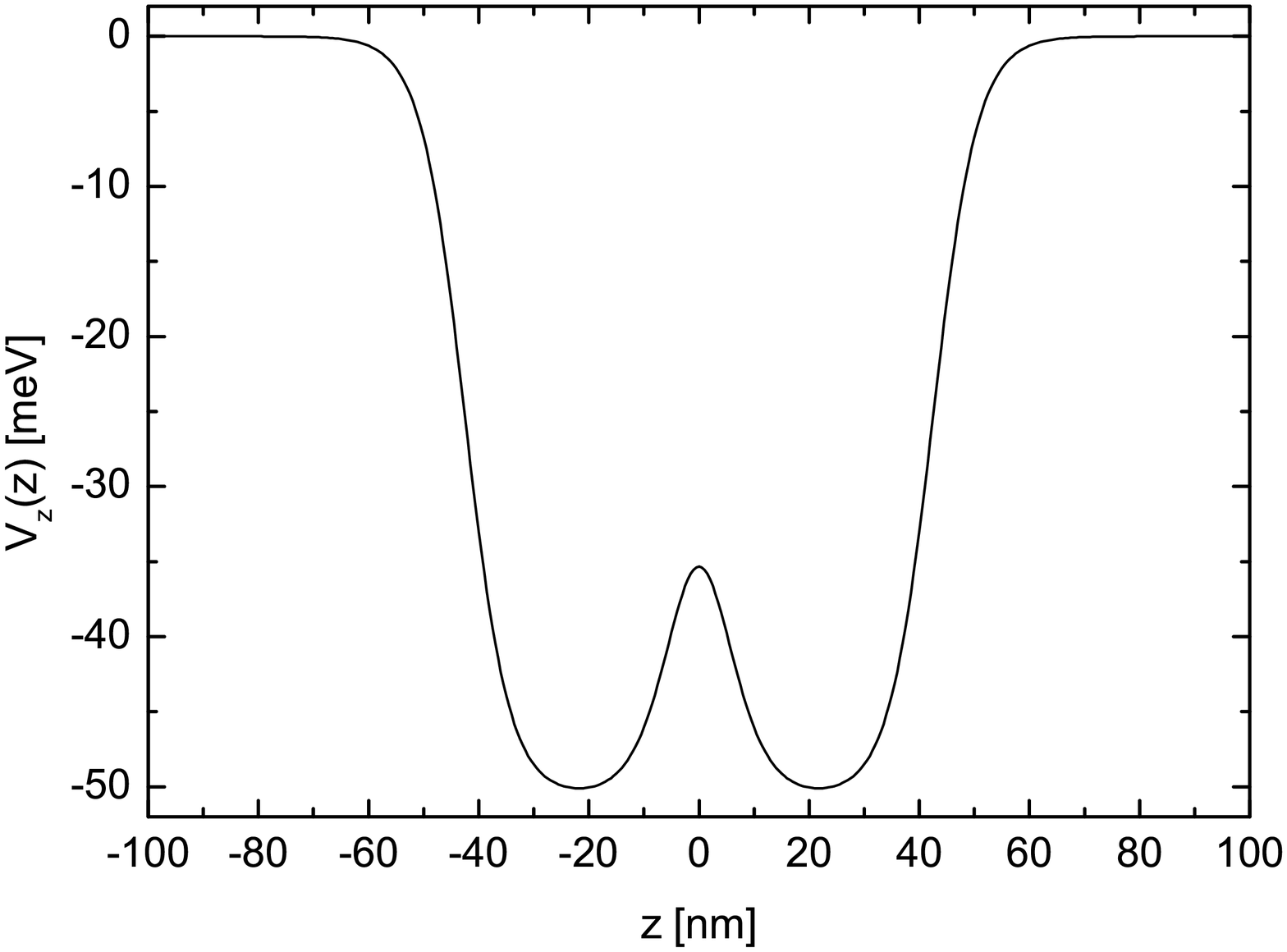}} &
\rotatebox{0}{\epsfxsize=55mm \epsfbox[0 15 565 573] {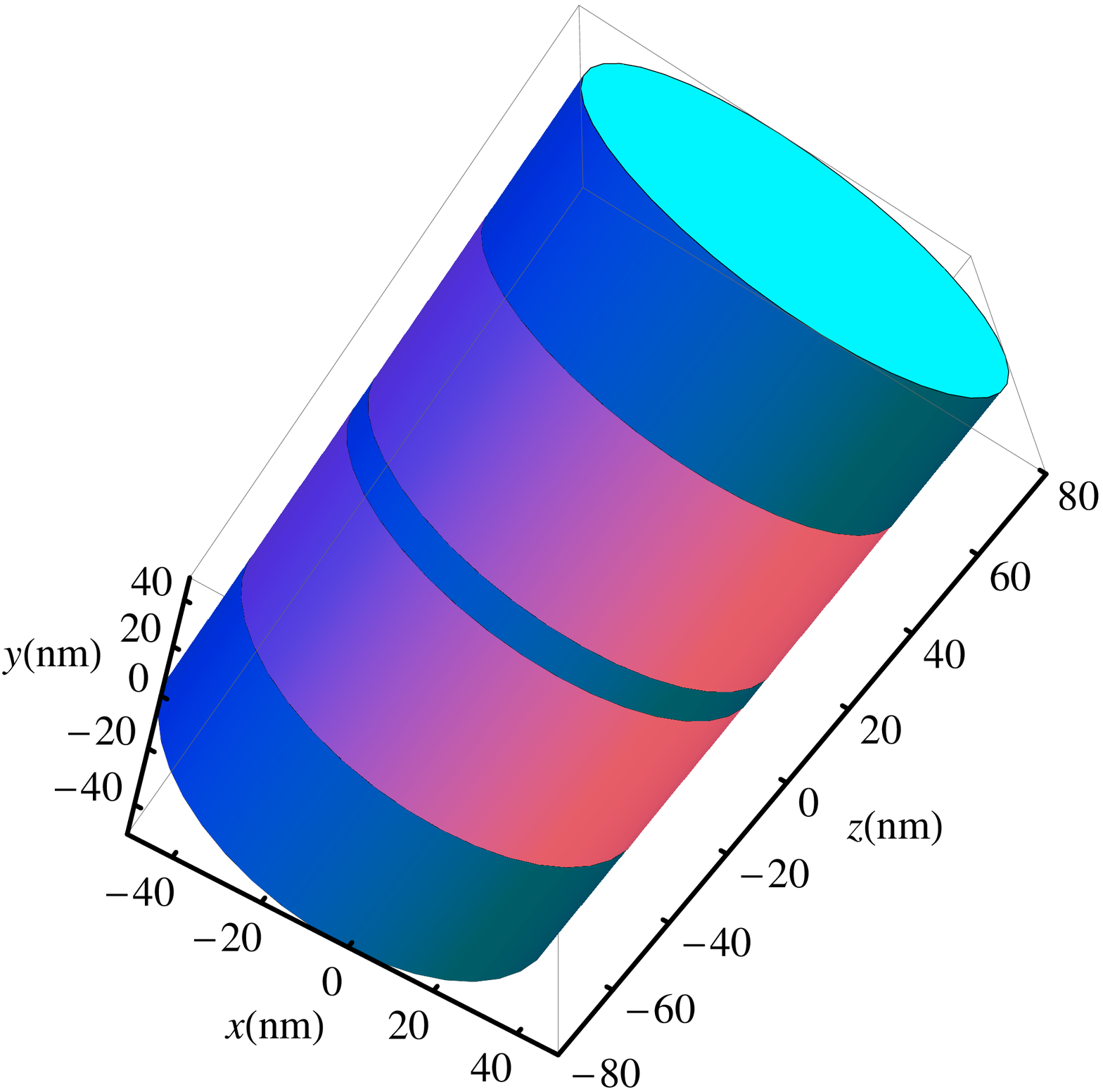}}
\end{tabular}
\caption{(a) The confinement potential along the $z$ axis. (b) Shape of the dots. The $z$ boundary is shown \mbox{for $V_z(z) = \frac{V_0}{2}$.}\label{potential}}
\end{figure*}

We model a confinement potential of two vertically stacked quantum dots in the form of an infinite circular quantum well in the $xy$ plane (corresponding to the cross-section of the wire) and two finite quantum wells along the $z$ axis (the growth axis of the wire). The zero of the energy scale is set to the degenerated top of heavy and light hole bands outside the dots. The total potential is $\hat{U} = (V_{xy}(\rho) + V_z(z)) \mathbf{I}$, where $\mathbf{I}$ is the unity matrix,
\begin{equation}
V_{xy}(\rho) = \begin{cases}
0 & \rho \leq R_{dot}\\
\infty & \rho > R_{dot}
\end{cases}
\end{equation} and
\begin{eqnarray}
V_z(z) &=& V_0 (V^{d}_z(z+z_{0}) + V^{d}_z(z-z_{0})),\nonumber\\
V^{d}_z(z) &=& - \frac{e^{\frac{z}{4}} \left(1 + e^{\frac{H_{d}}{8}}\right)^2}{e^{\frac{z}{4}} + e^{\frac{z + H_{d}}{4}} + e^{\frac{4 z + H_{d}}{8}} + e^{\frac{H_{d}}{8}}},\nonumber\\
z_{0} &=& \frac{1}{2} \left(H_{b} + H_{d}\right).
\label{confinement}\end{eqnarray}
In the equation above, the $V^{d}_z(z+z_0)$ part corresponds to the shape of the confinement of one of the dots and $V^{d}_z(z-z_0)$ part corresponds to the shape the confinement of the other one. The \mbox{$V_0 = 50$ meV} is the depth of the confinement. The \mbox{$H_{b} = 11$ nm} parameter describes the separation of the dots and the \mbox{$H_{d} = 40$} nm describes the width of the dots. The $V_z(z)$ potential is presented in \mbox{Fig.~\ref{potential}(a)} and the shape of the dots in \mbox{Fig.~\ref{potential}(b).} 

The adopted potential defining the system is symmetric in respect to reversing the nanowire $z$ axis. In any experimental realization, the potential for each dot would be slightly different. We have studied the impact of the asymmetry of the dots in Appendix \ref{AppD}.

\subsection{Kohn-Luttinger Hamiltonian}
We work in the effective mass approximation. The kinetic energy of holes is calculated using the 4-band Kohn-Luttinger Hamiltonian.\cite{chuang} The Hamiltonian for the $\langle100\rangle$ crystal orientation is given (in atomic units) by: \begin{eqnarray}
\hat{T}_{100} = &&\frac{1}{2}\left(\gamma_1 + \frac{5}{2} \gamma_2\right) k^2 \mathbf{I} - 2 \gamma_2 \left(k_x^2 J_x + k_y^2 J_y + k_z^2 J_z\right) \nonumber\\
&-& 4 \gamma_3 \left(k_x k_y J_{xy} + k_y k_z J_{yz} + k_z k_x J_{zx} \right),
\label{KL100}
\end{eqnarray} where $J_x, J_y, J_z$ are the spin matrices for spin $\frac{3}{2}$, $J_{ij} = \frac{1}{2} \left(J_i J_j + J_j J_i\right)$, $\gamma_{1}, \gamma_{2}, \gamma_{3}$ are the Luttinger parameters, and $k^2 = k_x^2 + k_y^2 + k_z^2$.\cite{KL111-1,KL111-2} To obtain the expression of the same Hamiltonian for $\langle111\rangle$ orientation, one should express the $(k_x,k_y,k_z)_{100}$ and $(J_x, J_y, J_z)_{100}$ vectors of the $\langle100\rangle$ orientation in the terms of $(k_x,k_y,k_z)_{111}$ and $(J_x, J_y, J_z)_{111}$ vectors of the $\langle111\rangle$ orientation, respectively (see Appendix \ref{AppA}).

If written in the basis $(HH\uparrow, LH\downarrow, LH\uparrow, HH\downarrow) = (|\frac{3}{2}, +\frac{3}{2}>, |\frac{3}{2}, -\frac{1}{2}>, |\frac{3}{2},+\frac{1}{2}>, |\frac{3}{2},-\frac{3}{2}>)$ the Hamiltonian for the $\langle111\rangle$ orientation has the following form \begin{equation}
\hat{T}= \left(
\begin{array}{cccc}
\hat{P}_{+} & \hat{R}_{s}+\hat{R}_{as} & -\hat{S}_{s}-\hat{S}_{as} & 0 \\
\hat{R}^{*}_{s}+\hat{R}^{*}_{as} & \hat{P}_{-} & 0 & \hat{S}_{s}+\hat{S}_{as} \\ 
-\hat{S}^{*}_{s}-\hat{S}^{*}_{as} & 0 & \hat{P}_{-} & \hat{R}_{s}+\hat{R}_{as} \\
0 & \hat{S}^{*}_{s}+\hat{S}^{*}_{as} & \hat{R}^{*}_{s}+\hat{R}^{*}_{as} & \hat{P}_{+}
\end{array}
\right),
\label{KL}
\end{equation}
where the operators used in this definition are listed in Table \ref{HKL_constants}.

\begin{table}
\begin{tabular}{|c|c|}
\hline $\hat{P}_{+} = \frac{(\gamma_{1} + \gamma_{3}) \hat{k}^{2}_{\perp} + (\gamma_{1} - 2 \gamma_{3}) \hat{k}^2_{z}}{2}$ & $\hat{P}_{-} = \frac{(\gamma_{1} - \gamma_{3}) \hat{k}^{2}_{\perp} + (\gamma_{1} + 2 \gamma_{3}) \hat{k}^2_{z}}{2}$\\
\hline $\hat{R}_{s} = -\frac{\sqrt{3}}{6} (\gamma_2 + 2 \gamma_3) \hat{k}^{2}_{-}$ & $\hat{R}_{as} = \frac{\sqrt{6}}{3} (\gamma_2 - \gamma_3)  \hat{k}_{+} \hat{k}_{z}$\\
\hline $\hat{S}_{s} = \frac{\sqrt{3}}{3} (2 \gamma_2 +\gamma_{3}) \hat{k}_{-} \hat{k}_{z}$ & $\hat{S}_{as} = - \frac{\sqrt{6}}{6} (\gamma_2 - \gamma_3) \hat{k}^{2}_{+}$\\
\hline $\hat{k}_{-} = \hat{k}_{x}-i\hat{k}_{y}$ & $\hat{k}_{+} = \hat{k}_{x}+i\hat{k}_{y}$\\
\hline $\hat{k}^{2}_{\perp} = \hat{k}^{2}_{x}+\hat{k}^{2}_{y}$ & \\
\hline
\end{tabular}
\caption{The operators used in the KL Hamiltonian definition.}
\label{HKL_constants}
\end{table}

The Kohn-Luttinger Hamiltonian can be divided into two parts:
\begin{eqnarray}
\hat{T} &=& \hat{T}_{s} + \hat{T}_{as}\nonumber,\\
\hat{T}_{s} &=& \left(
\begin{array}{cccc}
\hat{P}_{+} & \hat{R}_{s} & -\hat{S}_{s} & 0 \\
\hat{R}^{*}_{s} & \hat{P}_{-} & 0 & \hat{S}_{s} \\ 
-\hat{S}^{*}_{s} & 0 & \hat{P}_{-} & \hat{R}_{s} \\
0 & \hat{S}^{*}_{s} & \hat{R}^{*}_{s} & \hat{P}_{+}
\end{array}\right)\nonumber,\\
\hat{T}_{as} &=& \left(\begin{array}{cccc}
0 & \hat{R}_{as} & -\hat{S}_{as} & 0 \\
\hat{R}^{*}_{as} & 0 & 0 & \hat{S}_{as} \\ 
-\hat{S}^{*}_{as} & 0 & 0 & \hat{R}_{as} \\
0 & \hat{S}^{*}_{as} & \hat{R}^{*}_{as} & 0
\end{array}
\right).
\label{KL_div}
\end{eqnarray}

The $\hat{T}_{s}$ part is axially symmetric and hence its eigenstates have defined $z$-components of total angular momentum $J_{z} = J^{Bl}_{z} + J^{en}_{z}$, which is the sum of Bloch $J^{Bl}_{z}$ and envelope $J^{en}_{z}$ $z$-components. For this reason, the computation of the system described by $\hat{T}_{s}$ is much easier as the process for each one of $J_{z}$ subspaces can be done separately. The envelope eigenfunctions of $\hat{T}_{s}$ are 4-dimensional vector functions:
\begin{equation}
\Psi^{ax}_{(J_z,m)}(\vec{r}_{h}) =
\left(
\begin{array}{c}
\xi^{HH \uparrow}_{j_{h}} e^{i (J_z - 3/2) \phi} \\
\xi^{LH \downarrow}_{j_{h}} e^{i (J_z + 1/2) \phi} \\
\xi^{LH \uparrow}_{j_{h}} e^{i (J_z - 1/2) \phi} \\
\xi^{HH \downarrow}_{j_{h}} e^{i (J_z + 3/2) \phi}
\end{array}
\right).
\label{ho1peigen}
\end{equation}
Moreover, the non-axially-symmetric part $\hat{T}_{as}$ is relatively small: the constant in $\hat{R}_{s}$ is about 14 times greater than the one in $\hat{R}_{as}$, and the constant in $\hat{S}_{s}$ about 54 times greater than the one in $\hat{S}_{as}$.

The diagonal terms of the $\hat{T}_s+\hat{U}$ Hamiltonian:
\begin{eqnarray}
\hat{H}_{HH\uparrow} &=& \hat{H}_{HH\downarrow} = \hat{P}_{+} + \left(V_{xy}(\rho) + V_z(z)\right),\nonumber\\
\hat{H}_{LH\uparrow} &=& \hat{H}_{LH\downarrow} = \hat{P}_{-} + \left(V_{xy}(\rho) + V_z(z)\right),
\label{hole_1band}
\end{eqnarray}
have the corresponding envelope eigenfunctions:
\begin{eqnarray}
\psi^{HH}_{k,J^{en}_{z},n}(\vec{r}_{h}) &=& e^{i J^{en}_{z} \phi} \chi_{J^{en}_{z}}\left(\frac{\chi_{0}\left(k,J^{en}_{z}\right) \rho}{R}\right) Z^{HH}_{n}(z),\nonumber\\
\psi^{LH}_{k,J^{en}_{z},n}(\vec{r}_{h}) &=& e^{i J^{en}_{z} \phi} \chi_{J^{en}_{z}}\left(\frac{\chi_{0}(k,J^{en}_{z}) \rho}{R}\right) Z^{LH}_{n}(z), \label{hole_1band_eigen}
\end{eqnarray}
where $\chi_{J^{en}_{z}}$ is a Bessel function of the first kind of $J^{en}_{z}$-th order and $\chi_{0}\left(k,J^{en}_{z}\right)$ is the $k$-th zero of that function. The $n$ quantum numbers order the $Z^{HH}_{n}(z)$ functions by ascending energy and (separately) order the $Z^{LH}_{n}(z)$ functions in the same way.

Finally, the Hamiltonian for the axially-symmetric static system is
\begin{equation}
\hat{H}_s = \hat{T}_{s} + \hat{U} + \hat{H}_{B_z} + \hat{H}_{F_z},
\label{oneparticle_symm}
\end{equation}
where $\hat{H}_{B_z}$ and $\hat{H}_{F_z}$ are the magnetic field and electric field Hamiltonians (as defined below), respectively. As~the last two terms do not mix states with different $J_z$ quantum numbers, the eigenfunctions of Eq. (\ref{oneparticle_symm}) also have the form presented in Eq. (\ref{ho1peigen}).

\subsection{Electric and magnetic fields}
The Hamiltonian of the external electric field, applied along the growth $z$-axis, in atomic units, has the form of 
\begin{eqnarray}
\hat{H}_{F_z} &=& F_z z s(z) \nonumber\\
s(z) &=& \begin{cases}
1 & |z| \leq 70\,\text{nm}\\
\frac{100}{|z|} & |z| \geq 130\,\text{nm}\\
|s_p(z)| & 70\,\text{nm} \leq |z| \leq 130\,\text{nm},
\end{cases}
\end{eqnarray} where $F_z$ is the electric field amplitude and $s(z)$ is the shape function. The $s(z)$ has such character that it simulates an electric field that: I) is homogeneous in the area of the dots, II) is continuous everywhere up to second derivative and III) decreases as $\frac{1}{|z|}$ in area far from the system. The $s_p(z)$ is the simplest polynomial that meets the continuity assumptions.\cite{annotation1}

The Hamiltonian of the homogeneous magnetic field $\vec{B} = \left(0,0,B_z\right)$, in atomic units, is given by:
\begin{eqnarray}
\left(\hat{H}_{B_z}\right)_{11} &=& \frac{B_z}{2} \left(\left(\gamma_1 + \gamma_3\right) \left(J_z - \frac{3}{2}\right) + \frac{3 \kappa}{2}\right),\nonumber\\
\left(\hat{H}_{B_z}\right)_{22} &=& \frac{B_z}{2} \left(\left(\gamma_1 - \gamma_3\right) \left(J_z + \frac{1}{2}\right) - \frac{\kappa}{2}\right),\nonumber\\
\left(\hat{H}_{B_z}\right)_{33} &=& \frac{B_z}{2} \left(\left(\gamma_1 - \gamma_3\right) \left(J_z - \frac{1}{2}\right) + \frac{\kappa}{2}\right),\nonumber\\
\left(\hat{H}_{B_z}\right)_{44} &=& \frac{B_z}{2} \left(\left(\gamma_1 + \gamma_3\right) \left(J_z + \frac{3}{2}\right) - \frac{3 \kappa}{2}\right),\nonumber\\
\left(\hat{H}_{B_z}\right)_{ij} &=& 0,~i{\neq}j,
\end{eqnarray}
where $\kappa$ is the $g$-factor for heavy and light holes in the system. This is a model that was used in Ref. \onlinecite{climente_magnetic}, but with two changes: I) the inverted effective mass values for heavy holes in $xy$ plane $\gamma_1+\gamma_2$ and the light holes one $\gamma_1-\gamma_2$ for the $\langle100\rangle$ orientation were substituted by analogous values for the $\langle111\rangle$ system (\mbox{i.e. $\gamma_1+\gamma_3$} and $\gamma_1-\gamma_3$, respectively) and II) we omit the terms proportional to $B_z^{2}$ as they are very small for the range of magnetic field that was considered.\cite{annotation2} The $g$-factor for bulk InSb is equal to $15.6$, but in a system of this type the value is significantly quenched, i.e. $\kappa \in \left(0, 4\right)$.\cite{holes_exp} The Lande value, which does not take into account the influence of remote bands, is $4/3$. We decided to adopt a middle value of $\kappa=2.0$.

\subsection{Dresselhaus Hamiltonian}
In order to account for the mixing of the states with different spins, the Dresselhaus Hamiltonian was included. In the case of the $\langle100\rangle$ crystal orientation it has the form of
\begin{eqnarray}
\hat{H}^{100}_{D} =& &\frac{2}{\sqrt{3}} C_k (k_x\{J_x,J_y^2-J_z^2\} + \text{c.p.})\nonumber\\
&+& b_{41}(\{k_x,k_y^2-k_z^2\}J_x + \text{c.p.})\nonumber\\
&+& b_{42}(\{k_x,k_y^2-k_z^2\}J_x^3 + \text{c.p.})\nonumber\\
&+& b_{51}(\{k_x,k_y^2+k_z^2\}\{J_x,J_y^2-J_z^2\} + \text{c.p.})\nonumber\\
&+& b_{52}(k_x^3\{J_x,J_y^2-J_z^2\} + \text{c.p.}),
\label{Dr_100}
\end{eqnarray}
where $C_k$, $b_{41}$, $b_{42}$, $b_{51}$, $b_{52}$ are material parameters, \mbox{$\{A,B\}=\frac{1}{2}(AB+BA)$} and c.p. stands for cyclic permutations of the preceding terms.\cite{climente_dresselhaus,winkler} The procedure for obtaining the Hamiltonian for the $\langle111\rangle$ orientation is the same as for the Kohn-Luttinger Hamiltonian (see Appendix \ref{AppA}). It leads to the following result:
\begin{eqnarray}
\hat{H}_{D} &=& \left(
\begin{array}{cccc}
\hat{O}_{1} & \hat{O}_{3} & \hat{O}_{2} & \hat{O}_{4} \\
\hat{O}_{3}^{+} & -a\hat{O}_{1} & \hat{O}_{5} & \hat{O}_{2} \\ 
\hat{O}_{2}^{+} & \hat{O}_{5}^{+} & a\hat{O}_{1} & -\hat{O}_{3} \\
\hat{O}_{4}^{+} & \hat{O}_{2}^{+} & -\hat{O}_{3}^{+} & -\hat{O}_{1}
\end{array}\right),
\label{Dr_111}
\end{eqnarray}
where the element operators are defined as follows:
\begin{eqnarray}
\hat{O}_{1} &=& -c_{1} (i \hat{k}_{-})^3 + c_{1} (i \hat{k}_{+})^3,\nonumber\\
\hat{O}_{2} &=& -\frac{C_k}{\sqrt{3}} (i \hat{k}_{-}) + c_{2} \hat{k}^{2}_{\perp} (i \hat{k}_{-}) - i c_{3} \hat{k}_z (i \hat{k}_{+})^2 + c_{4} \hat{k}_z^2 (i \hat{k}_{-}),\nonumber\\
\hat{O}_{3} &=& \frac{C_k}{\sqrt{6}} (i \hat{k}_{+}) + c_{5} \hat{k}^{2}_{\perp} (i \hat{k}_{+}) - i \sqrt{3} c_{6} \hat{k}_z (i \hat{k}_{-})^{2},\nonumber\\
\hat{O}_{4} &=& c_{6} (i \hat{k}_{-})^3 - c_{8} (i \hat{k}_{+})^3 - i c_{13} \hat{k}_z \hat{k}^{2}_{\perp}\nonumber\\
& &- i \sqrt{2} C_k \hat{k}_z - i c_{9} \hat{k}_z^3,\nonumber\\
\hat{O}_{5} &=& -C_k (i \hat{k}_{+}) - c_{10} \hat{k}^{2}_{\perp} (i \hat{k}_{+}) + i c_{11} \hat{k}_z (i \hat{k}_{-})^2 \nonumber\\
& & - c_{12} \hat{k}_z^2 (i \hat{k}_{+}),
\label{Dr_111_oper}
\end{eqnarray}
where constants $a$ and $c_1$ to $c_{13}$ are defined in the terms of $b_{41}$, $b_{42}$, $b_{51}$, $b_{52}$ (see Table \ref{Dress_constants}). Please note, that if the angular dependencies of ${\mid}L\rangle$ and ${\mid}R\rangle$ states are $e^{i l_L \phi}$ and $e^{i l_R \phi}$, respectively, then the ${\langle}L{\mid}\hat{k}_{-}{\mid}R{\rangle}$ matrix element is nonzero only for $l_R=l_L+1$, the ${\langle}L{\mid}\hat{k}_{+}{\mid}R{\rangle}$ matrix element is nonzero only for $l_R=l_L-1$, the ${\langle}L{\mid}\hat{k}^{2}_{\perp}{\mid}R{\rangle}$ and the ${\langle}L{\mid}\hat{k}_{z}{\mid}R{\rangle}$ matrix elements are nonzero only for $l_R=l_L$. This leads to a significant simplification of the Hamiltonian, see Appendix \ref{AppB}.
\begin{table}
\begin{tabular}{|c|c|}
\hline $c_1 = \frac{12 b_{41}+23 b_{42}}{16 \sqrt{6}}$ & $c_2 = -\frac{4 b_{41}+9 b_{42}+4 (b_{51}+b_{52})}{16}$\\
\hline $c_3 = \frac{4 b_{41}+9 b_{42}-4 b_{51}+4 b_{52}}{8 \sqrt{2}}$ & $c_4 = \frac{4 b_{41}+9 b_{42}-2 b_{52}}{4}$\\
\hline $c_5 = \frac{-b_{42}+b_{51}+b_{52}}{4 \sqrt{2}}$ & $c_6 = \frac{b_{42}+b_{51}-b_{52}}{4 \sqrt{3}}$\\
\hline $c_7 = \frac{2 b_{42}+b_{52}}{2 \sqrt{2}}$ & $c_8 = \frac{b_{42}-b_{51}+b_{52}}{4 \sqrt{3}}$\\
\hline $c_9 = \frac{2 b_{51}+b_{52}}{\sqrt{6}}$ & $c_{10} = \frac{-4 b_{41}-7 b_{42}+6 (b_{51}+b_{52})}{8 \sqrt{3}}$\\
\hline $c_{11} = \frac{4 b_{41}+7 b_{42}+6 b_{51}-6 b_{52}}{4 \sqrt{6}}$ & $c_{12} = \frac{4 b_{41}+7 b_{42}+3 b_{52}}{2 \sqrt{3}}$\\
\hline $c_{13} = \sqrt{\frac{3}{2}} b_{52}$ & $ a = \frac{4 b_{41} + 13 b_{42}}{12 b_{41} + 23 b_{42}}$ \\
\hline
\end{tabular}
\caption{The material constants used for Dresselhaus Hamiltonian definition.}
\label{Dress_constants}
\end{table}

\subsection{Computational method}
Our computational method consists of several separable steps. At the beginning, the one-band hole Hamiltonian eigenequations Eq.\,(\ref{hole_1band}) are solved. The $Z^{HH}_{n}(z)$ and $Z^{LH}_{n}(z)$ functions in Eq.\,(\ref{hole_1band_eigen}) are determined by direct diagonalisation on one-dimensional mesh with mesh spacing $\Delta_{z} = 0.5$ nm and computation box of \mbox{$z\in(-200~\text{nm},200~\text{nm})$.} Afterwards, the $\Psi_{\left(J_z,m\right)}^{ax}(\vec{r})$ eigenfunctions of the axially-symmetric Hamiltonian Eq.\,(\ref{oneparticle_symm}) are obtained in a base constructed by taking functions of type as in formula Eq.\,(\ref{hole_1band_eigen}) with $k \in \{1,8\}$, $n \in \{1,32\}$, and $J_z \in \{-\frac{13}{2},...,\frac{13}{2}\}$. As it was mentioned before, each $J_z$ defines a separable subspace and the $m$ quantum number sorts the eigenfunctions of each subspace in the order of ascending energy.

The next step is to include the non axial part into the calculation. This part consists of the small non-axial terms in Kohn-Luttinger Hamiltonian and of the Dresselhaus Hamiltonian:
\begin{equation}
\hat{H}_{as} = \hat{T}_{as}+\hat{H}_{D}.
\label{Has}
\end{equation}
This operation is made in a basis consisting of selected set of lowest-lying $\hat{H}_{s}$ eigenstates:
\begin{equation}
\Psi_{j}(\vec{r}) = \sum_{\left(J_z,m\right) \in \Omega_{N}} d^{N}_{j,\left(J_z,m\right)} \Psi_{\left(J_z,m\right)}^{ax}(\vec{r}),
\label{Psi_as}
\end{equation}
where $\Omega_{N}$ is the basis for the non-axially-symmetric calculation and $d^{N}$ is the projection of the $j$-th non-axial state onto individual $\left(J_z,m\right)$ basis state.

\subsection{Evolution}
The evolution simulation of the system is done with the Runge-Kutta method of the $4$-th order with the time step of ${\Delta}_t = 0.2$ fs. The basis for the evolution is the set of states obtained in the non-axially-symmetric calculation, without the external electric field:
\begin{equation}
\Phi(\vec{r},t) = \sum_{j \in \Omega_T} d^{T}_{j}(t) \Psi_{j}(\vec{r};F_z = 0),
\label{time_evolution_basis}
\end{equation}
where $\Omega_{T}$ is the evolution basis and $d^{T}_{j}$ is the projection of the time-dependent state onto individual $j$-th basis state (in order of ascending energy). The specific algorithm for the evolution of $d^{T}_{j}$ projections is given below:
\begin{eqnarray}
d^{T}_{j}(t_{n+1}) &=& \frac{1}{3} \left(\left[x_1\right]_{j} + 2 \left[x_2\right]_{j} + \left[x_3\right]_{j} \right.\nonumber\\
&&+\left.\left[x_4\right]_{j} - d^{T}_{j}\left(t_n\right)\right),\nonumber\\
\left[x_1\right]_{j} &=& d^{T}_{j}\left(t_n\right) - \frac{i {\Delta}_t}{2} \langle \Psi^{F_z=0}_{j} \mid \hat{H}_{t_n} \mid \Phi_{t_n} \rangle, \nonumber\\
\left[x_2\right]_{j} &=& d^{T}_{j}\left(t_n\right) - \frac{i {\Delta}_t}{2} \langle \Psi^{F_z=0}_{j} \mid \hat{H}_{t_{(n+1/2)}} \mid x_1 \rangle, \nonumber\\
\left[x_3\right]_{j} &=& d^{T}_{j}\left(t_n\right) - i {\Delta}_t \langle \Psi^{F_z=0}_{j} \mid \hat{H}_{t_{(n+1/2)}} \mid x_2 \rangle, \nonumber\\
\left[x_4\right]_{j} &=& - \frac{i {\Delta}_t}{2} \langle \Psi^{F_z=0}_{j} \mid \hat{H}_{t_{n+1}} \mid x_3 \rangle,
\end{eqnarray}
where $j\in\Omega_T$.

Please note that all results for the evolution are presented in \textit{local $F_z$} basis, and not in the $F_z=0$ basis, for ease of interpretation. The projections $d^{T}_{j}(t)$, obtained as have been explained above, are recalculated to represent the projections as if the levels $\Psi_{j}(\vec{r})$ at each given $F_z$ were the basis states instead of the $\Psi_{j}(\vec{r};F_z = 0)$ states. In this way, one can refer to these projections as corresponding to the hole energy spectrum in each point of the $F_z$ axis.

\subsection{Parametrisation}
The values of material constants for the Dresselhaus Hamiltonian, i.e. \mbox{$C_k = -0.82$ meV nm}, \mbox{$b_{41} = -934.8$ meV nm\textsuperscript{3}}, \mbox{$b_{42} = 41.73$ meV nm\textsuperscript{3}}, \mbox{$b_{51} =$} \mbox{$13.91$ meV nm\textsuperscript{3}}, \mbox{$b_{52} = -27.82$ meV nm\textsuperscript{3}} are taken from Ref. \onlinecite{winkler}. All the other material parameters were taken from the work of Vurgaftman \textit{et al.}\cite{bible} Luttinger parameters for InSb are \mbox{$\gamma_{1} =  34.8$}, \mbox{$\gamma_{2} = 15.5$}, \mbox{$\gamma_{3} = 16.5$}.

\section{Results}

\subsection{Time-independent system}

\begin{figure*}[!ht]
\makebox[\textwidth][c]{\begin{tabular}{cc}
(a) & (b)\\
\rotatebox{0}{\epsfxsize=70mm \epsfbox[47 30 705 505] {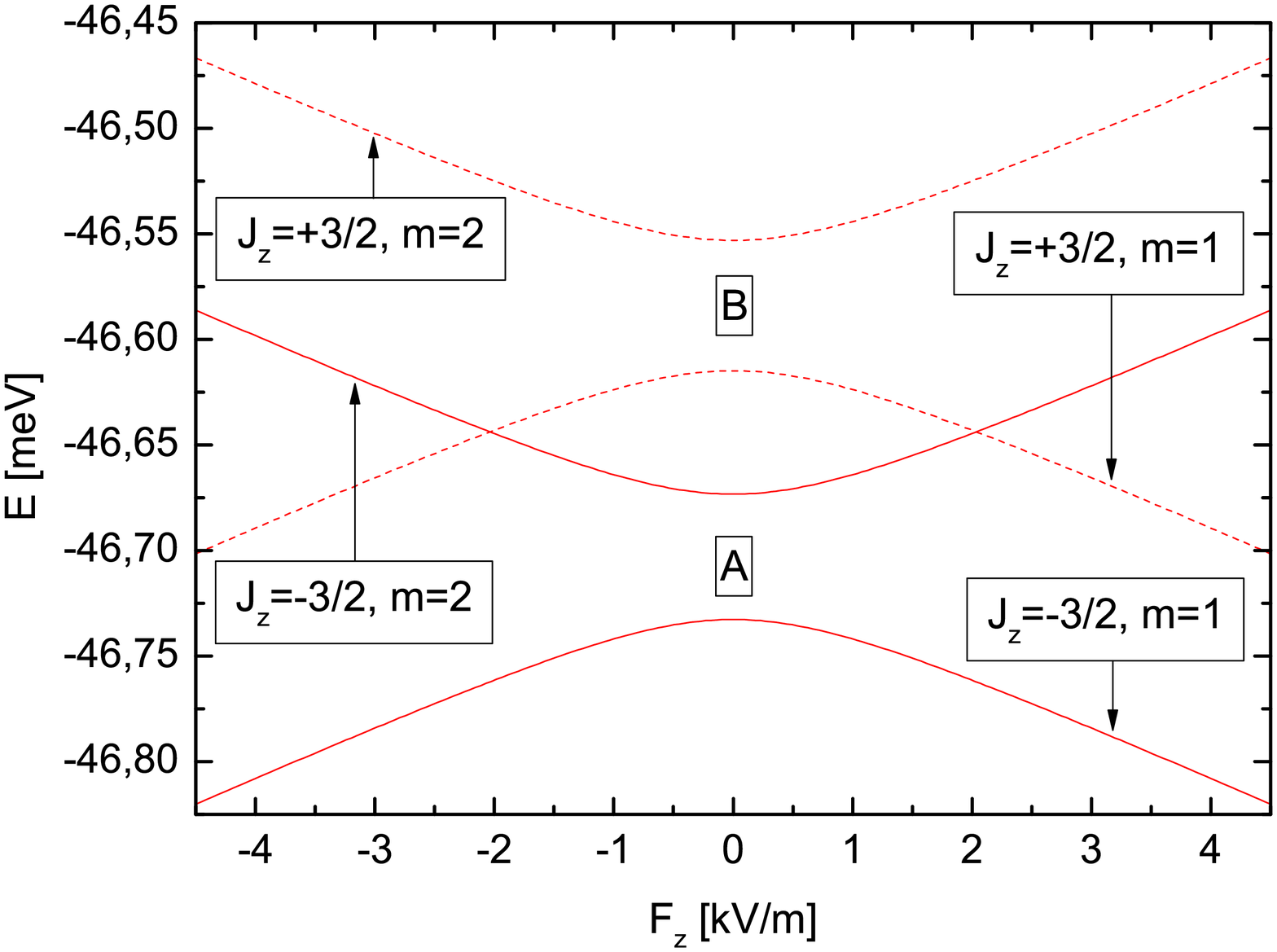}} &
\rotatebox{0}{\epsfxsize=70mm \epsfbox[43 65 700 540] {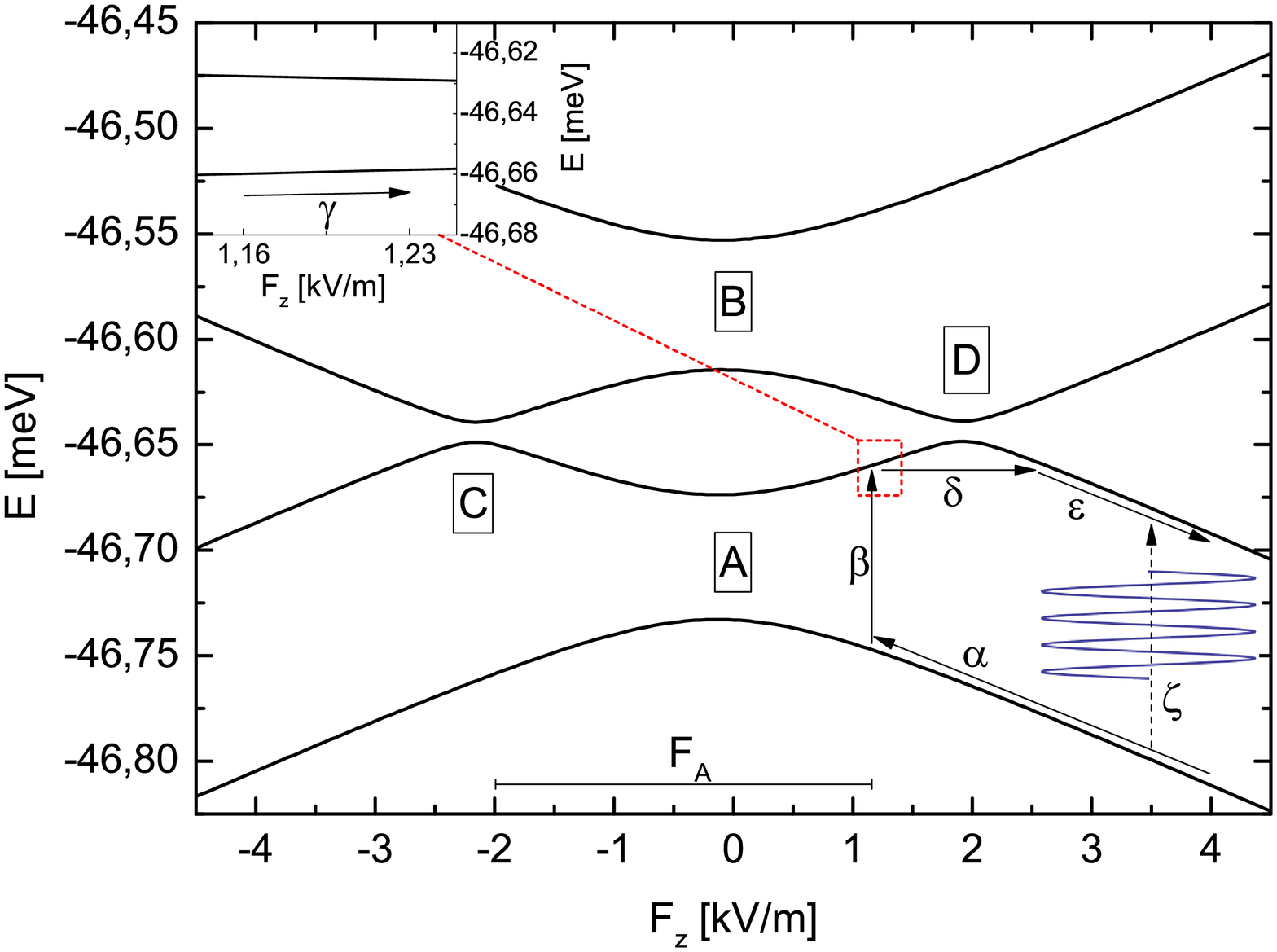}}
\end{tabular}}
\caption{(a) Hole energy spectrum of the axially symmetric system. (b) Hole energy spectrum of the system with non-axially-symmetric terms included and the scheme of the hole transfer. Arrows indicate the succesive steps of the process. $F_A$ is the field range used for the Landau-Zener transfer in the $\beta$ step. The dashed $\zeta$ arrow denotes an alternative EDSR approach (see Sec. \ref{comparison}). The inset is a fragment of the spectrum in magnification.\label{Hs_spectrum}}\end{figure*}

We begin by studying the energy spectrum with a static electric field $F_z$ applied. The energy spectrum of the hole system for the axially symmetric Hamiltonian $H_s$ [see Eq.~(\ref{oneparticle_symm})] is presented in \mbox{Fig.~\ref{Hs_spectrum}(a)}. The set of levels with the lowest energies has the following elements: ($J_z=-\frac{3}{2}$,$m=1$), ($J_z=-\frac{3}{2}$,$m=2$), ($J_z=\frac{3}{2}$,$m=1$), and ($J_z=\frac{3}{2}$,$m=2$). This set of levels is separated energetically from the next ones for any $F_z$ in the considered range by about $0.53$ meV. The characteristics of the four lowest-lying $\hat{H}_s$ eigenstates are given in Table \ref{Level_characteristics}. In every case, the dominating valence band is the one with the lowest $\left|J^{en}_{z}\right|$ value (that equals $0$ for the first four levels), and, in each case, it is one of the heavy hole bands.

\begin{table}
\begin{tabular}{|c|c|c|}
\hline Level & \begin{tabular}[c]{@{}c@{}}Dominating\\valence band\end{tabular} & set of $J^{en}_{z}$ values\\
\hline ($J_z=-\frac{3}{2}$,$m=1$) & $HH\downarrow$ & $(-3,-1,-2,0)$ \\
\hline ($J_z=-\frac{3}{2}$,$m=2$) & $HH\downarrow$ & $(-3,-1,-2,0)$ \\
\hline ($J_z=\frac{3}{2}$,$m=1$) & $HH\uparrow$ & $(0,2,1,3)$ \\
\hline ($J_z=\frac{3}{2}$,$m=2$) & $HH\uparrow$ & $(0,2,1,3)$ \\
\hline
\end{tabular}
\caption{The characteristics of the four lowest-lying $\hat{H}_s$ eigenstates.}
\label{Level_characteristics}
\end{table}

The main two features of this spectrum are the two avoided crossings: the one of the two $J_z=-\frac{3}{2}$ levels [marked as A in \mbox{Fig.~\ref{Hs_spectrum}(a)}] and the one of the two $J_z=\frac{3}{2}$ levels [marked as B in \mbox{Fig.~\ref{Hs_spectrum}(a)}]. The mentioned avoided crossings occur due to the tunneling coupling between the dots. At $F_z=0$ the confinement potential of the system is symmetric in respect to $z=0$, hence the eigenfunctions are equally distributed between both dots. For $F_z>>0$, away from the crossing, the hole is localized in the energetically preferable $z<0$ dot in the ground level of each $J_z$ subspace (i.e. levels with $m=1$) and in the energetically impreferable $z>0$ dot in the excited levels (i.e. levels with $m=2$). The situation is reversed for $F_z<<0$.

In the case of the non-axially-symmetric calculation, due to computational constraints, we are interested in the lowest-lying states only. Because of the energy separation of about $0.53$ meV, the first four energy levels, shown in \mbox{Fig.~\ref{Hs_spectrum}(a)}, create a natural basis for this calculation. Thus we define the basis $\Omega_N$ in \mbox{Eq. (\ref{Psi_as})} as the set of levels listed in Table \ref{Level_characteristics}.

The hole energy spectrum for the total Hamiltonian is presented in \mbox{Fig.~\ref{Hs_spectrum}(b)}. In comparison to the axial one [see \mbox{Fig.~\ref{Hs_spectrum}(a)}] the only important difference is the formation of two additional smaller avoided crossings, marked as C and D in \mbox{Fig.~\ref{Hs_spectrum}(b)}. These anticrossings correspond to the mixing of the ($J_z=-\frac{3}{2}$,$m=1$), that is the spin-down state, and ($J_z=\frac{3}{2}$,$m=1$) state, that is the spin-up one. Apart from that, the energy shifts are very small, and the \mbox{Fig.~\ref{Hs_spectrum}(b)} spectrum is nearly the same as the \mbox{Fig.~\ref{Hs_spectrum}(a)} one.

\begin{figure}[!ht]
\epsfxsize=70mm \epsfbox[0 0 670 610]{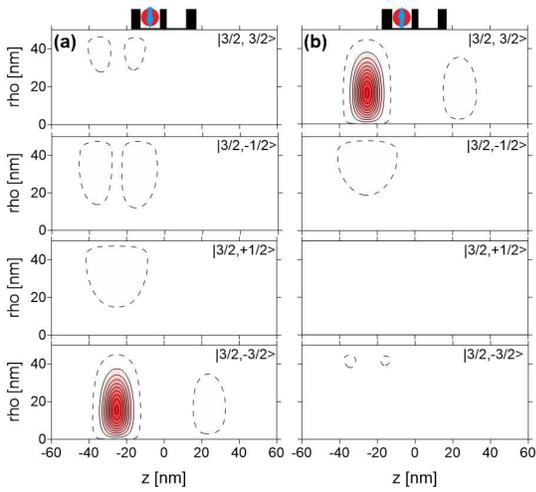}
\caption{$P_{j,k}(\rho,z)$ -- the hole one-band probability densities integrated over the $\phi$ coordinate -- for $F_z=4$ kV/m: (a) for the ground state, (b) for the first excited level. \label{probability_densities}}\end{figure}

The one-band probability densities, integrated over the~$\phi$ coordinate:
\begin{eqnarray}
P_{j,k}(\rho,z) &=& \rho \int_{0}^{2 \pi} \Psi_{j}^{*}(\rho,\phi,z) \mathbf{I}_k \Psi_{j}(\rho,\phi,z) d\phi \nonumber\\
\mathbf{I}_k &=& \left(
\begin{array}{cccc}
\delta_{1,k} & 0 & 0 & 0 \\
0 & \delta_{2,k} & 0 & 0 \\ 
0 & 0 & \delta_{3,k} & 0 \\
0 & 0 & 0 & \delta_{4,k}
\end{array}\right),
\end{eqnarray}
are shown for $F_z=4$ kV/m in Fig.~\ref{probability_densities}(a) for the ground state, and in Fig.~\ref{probability_densities}(b) for the first excited level, respectively. The ground state of the hole is localized in the $z<0$ dot and is strongly dominated by the $HH\downarrow$ band. The first excited state is localized in the same dot and is strongly dominated by the $HH\uparrow$ band. This corresponds to our idea to flip the spin of a hole by transferring it from the ground state to the first excited level. This would result in reversing the state from being $HH\downarrow$ dominated to being $HH\uparrow$ dominated, while remaining in the same $z<0$ dot. In order to do so, we plan to employ the Landau-Zener transitions of A and D avoided crossings, see Fig.~\ref{Hs_spectrum}(b).

\subsection{Evolution}

The idea of the spin flip is presented in Fig.~\ref{Hs_spectrum}(b). The initial state of the simulation is the time-independent ground state at \mbox{$F_z=4$ kV/m} and the intended final state is the time-independent first excited state at the same electric field. The transfer is planned to be made in five steps. It should start with tuning the electric field to the $F_z > 0$ side of the larger avoided crossing in such a way that the time-dependent state would remain equal to the time-independent ground level [arrow~$\alpha$ in Fig.~\ref{Hs_spectrum}(b)]. The second step consists of using this anticrossing to transfer the time-dependent hole state to the first excited level at the same electric field [arrow~$\beta$ in Fig.~\ref{Hs_spectrum}(b)]. The third stage is the drive of the hole from the tunneling-generated anticrossing to the \textit{lower $F_z$} side of the smaller spin-mixing one [arrow~$\gamma$ in the inset of	 Fig.~\ref{Hs_spectrum}(b)]. After that, the hole state should be transferred across the smaller avoided crossing without the \textit{leak} of the time-dependent state to another time-independent levels [arrow~$\delta$ in Fig.~\ref{Hs_spectrum}(b)]. It is here that the spin flip takes place. The final step is the drive of the system to \mbox{$F_z=4$ kV/m} [arrow~$\epsilon$ in Fig.~\ref{Hs_spectrum}(b)].

\begin{figure}[!ht]
\epsfxsize=70mm \epsfbox[65 35 750 540] {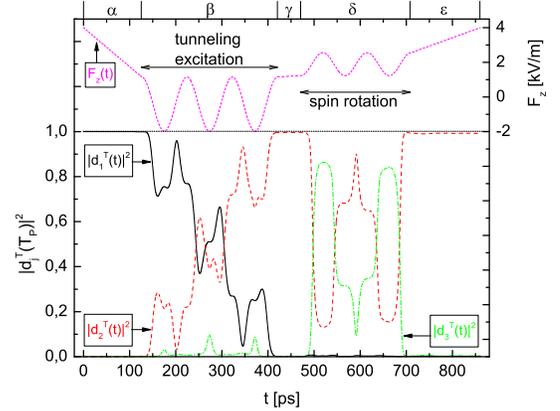}
\caption{The evolution process for the complete transition. Upper part, right axis: The driving field $F_z(t)$ function. Lower part, left axis: The~$\left|d_j^T(t)\right|^2$ projections for the time-dependent state $\Phi(\vec{r},t)$ on the $j$-th time-independent state \mbox{$\Psi_{j}(\vec{r})$}, respectively. For $j=4$ the projection is not shown as it is negligible at every moment of the evolution. \label{transfer_total}}
\end{figure}

Each of $\{\alpha, \beta, \gamma, \delta, \epsilon\}$ stages was optimized separately for transfer efficiency in terms of relevant parameters (see~Appendix \ref{AppC}) and the total $F_z(t)$ driving impulse was constructed by joining all the parts together. The evolution of the total transfer is presented in Fig. \ref{transfer_total}. The initial state is equal to the time-independent ground state \mbox{$\Psi_{1}(\vec{r})$}, that is $d_1^T(0)=1$. After the evolution, the evolving state ends in the first excited state \mbox{$\Psi_{2}(\vec{r})$} with $|d_2^T(T_P)|^2>0.99$. This means that the system started and ended evolution in the same electric field. Both the initial and the final states are localized in the same dot. However, the system started evolution in a state dominated by the $HH\downarrow$ spin state and it ended it in a state dominated by the opposite $HH\uparrow$ spin state. This proposition of the process that reverses the spin state of the hole is the main result of this work.

\subsection{Comparison with the EDSR of uncoupled levels\label{comparison}}

\begin{figure}[ht!]
\rotatebox{0}{\epsfxsize=70mm \epsfbox[65 37 682 542] {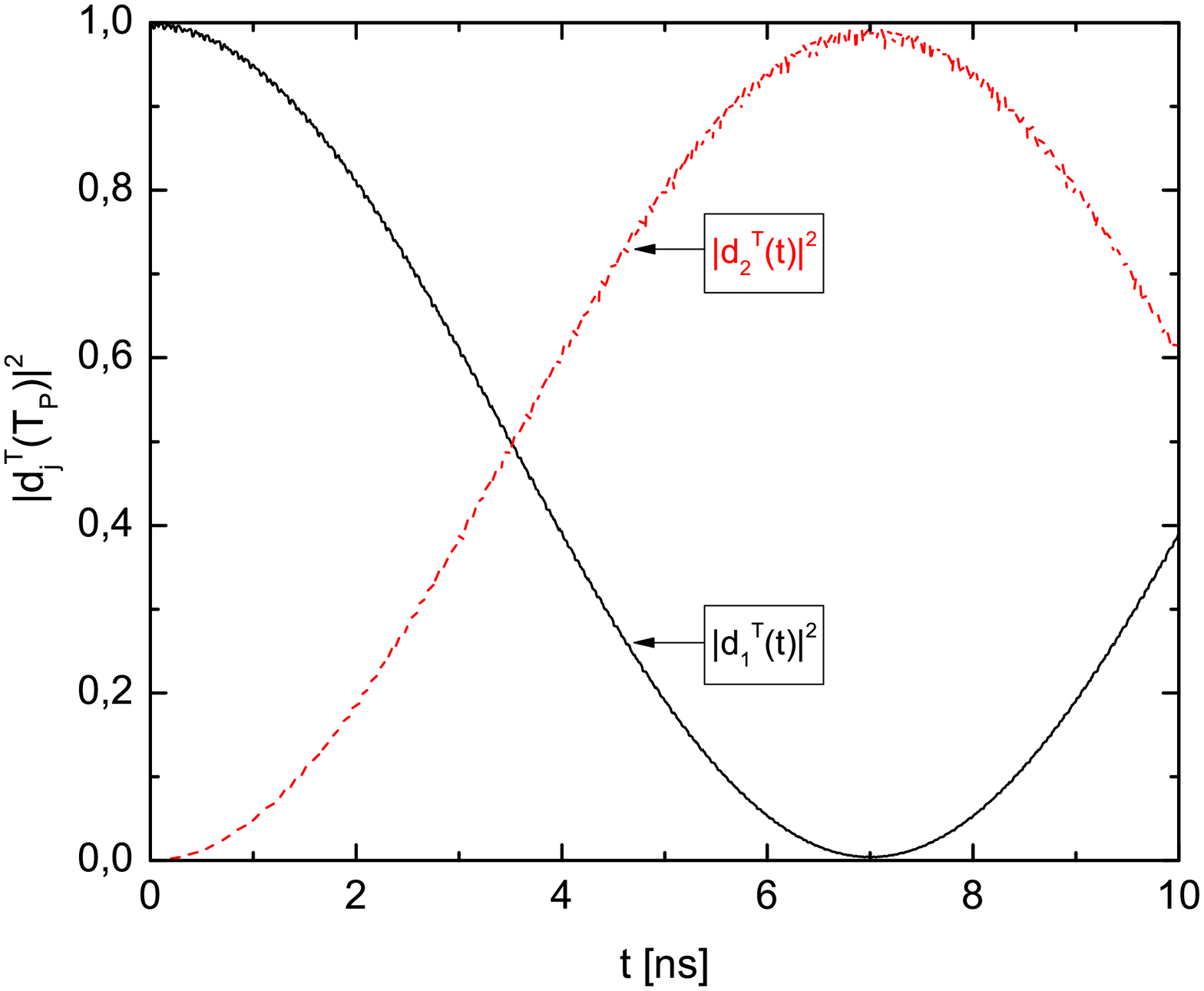}}
\caption{The evolution process for the EDSR of uncoupled levels. The~$\left|d_j^T(t)\right|^2$ projections for the time-dependent state $\Phi(\vec{r},t)$ on the $j$-th time-independent state \mbox{$\Psi_{j}(\vec{r})$}, respectively. For $j\in\lbrace3,4\rbrace$ the projections are not shown as they are negligible at every moment of the evolution.\label{EDSR}}
\end{figure}
The Landau-Zener type of spin-flip is an alternative for the EDSR one. In case of the latter the levels involved in transition are not engaged in an avoided crossing. In order to make the comparison between these two mechanisms, the transition $\zeta$ in Fig. \ref{Hs_spectrum}(b) was calculated. The frequency of the driving signal
\begin{equation}
F_z(t) = F_{o} + A \cos\left(\omega_c t\right)
\end{equation}
is tuned to resonance with the energy difference between the final and the initial level: $\omega_c = E_{f} - E_{i} = 2 \pi \cdot 28.8$ GHz for the offset $F_{o} = 3.5$ kV/m. The amplitude of EDSR signal is $A = 0.95	$ kV/m. The results are presented in Fig. \ref{EDSR}. The evolving state starts in the ground state of the time-independent system and then typical Rabi oscillations begin. The evolving state occupies the first excited state after $t \sim 7$ ns, which corresponds to about $200$ periods of the $F_z$ signal function. Please note that this time is about one order of magnitude larger that the time of the total evolution for the $\lbrace\alpha,\beta,\gamma,\delta,\epsilon\rbrace$ scheme.

\section{Noise}

\subsection{The models of the noise}

The evolution simulation assumes total control of the driving electric field $F_z(t)$. In an experiment, such a precise control is impossible. The impact of the \textit{power-law noise} on the effectiveness of the transfer is studied by implementation of a simple random telegraph noise model (RNT), as described in Ref. \onlinecite{RTN}. According to the model, the actual time dependence of the electric field is given by:
\begin{equation}
F_z^{e}(t) = F_z(t) + F_N G_i(t,f_c),
\end{equation}
where $F_z(t)$ is the non-distorted electric field drive, $F_N$ is the jump amplitude of RTN,
\begin{equation}
G_i(t,f_c) = C_{sgn} (-1)^{\sum_j \Theta(t-t_{i,j})}
\label{random_jumps}
\end{equation}
is the electric distortion of the RTN, $\Theta$ stands for the Heaviside step function and the time of the $j$-th jump is defined as follows:
\begin{equation}
t_{i,j} = - \frac{1}{f_c} \sum_{n=1}^j \ln p_{i,j}.
\label{random_jump_time}
\end{equation}
In Eqs. (\ref{random_jumps}) and (\ref{random_jump_time}) the $i$ variable is the current iteration of random generation of a set of $p_{i,j}$  numbers from a uniform distribution over $[0,1]$ range and the sign of the first jump $C_{sgn}$ is determined randomly. The characteristic frequency $f_c$ is related to the average number of jumps occurring during the evolution time $T_P$:
\begin{equation}
N_{avg} = T_P f_c.
\end{equation}

It was shown that this model simulates the \textit{power-law} noise well for a sufficiently high amount of jumps per one evolution ($f_c > 1$ GHz).\cite{RTN} On the opposite end of the scale, where the frequency of the noise change is low, e.g. if $90\%$ of cases have no jumps ($f_c < 100$ MHz), a different model was adopted. The little variability of the noise signal during the evolution can be simulated by adopting a static shift in electric field:
\begin{equation}
F_z^{e}(t) = F_z(t) \pm F_N,
\end{equation}
where the sign of the shift is determined randomly.

\subsection{The noise simulation}

\begin{figure}[!ht]
\epsfxsize=70mm \epsfbox[0 10 525 505]{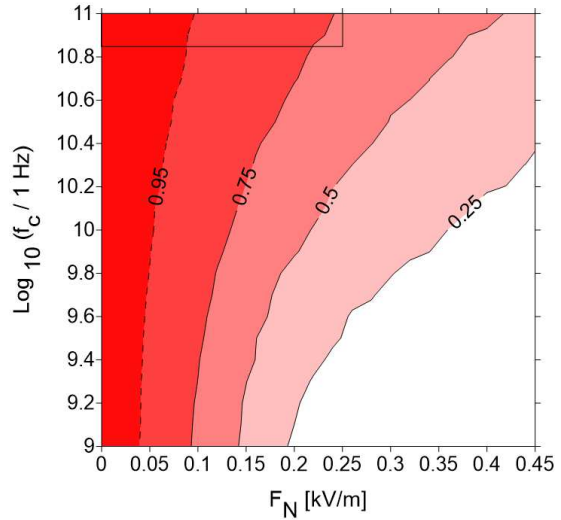}
\caption{The final $\left|d_2^T\right|^2$ projection for the RTN $F_z$ noise simulation (the fast-variable noise regime). $F_N$ is the amplitude of the noise and $f_c$ is the characteristic frequency of the noise. The square in the upper left corner marks the region where an analog of the motionary narrowing occurs (see Appendix \ref{AppE}).\label{noise}}
\end{figure}

The results for the RTN noise model are presented in Fig.~\ref{noise}. The final efficiency was averaged over  $1000$ simulations for each pair of $F_N$ and $f_c$ values. Keeping transfer efficiency equal, an increase in $f_c$ leads to an increase in $F_N$. The mechanism of the observed effect is very similar to the one responsible for the motional narrowing effect in magnetic resonance (see e.g. Ref. \onlinecite{motional} pp. 212-213 and Appendix \ref{AppE}). If the noise changes very quickly, then the system does not adapt to each individual shift value. The mean value of the noise shift is equal to zero, and so the overall effect of the noise is diminished in comparison to the noise with lower $f_c$.

\begin{figure}[!ht]
\epsfxsize=70mm \epsfbox[65 30 685 540]{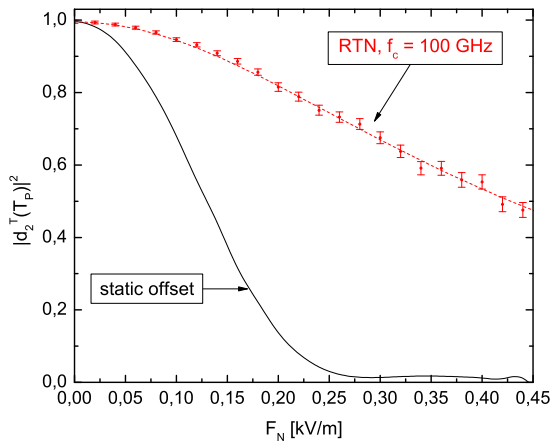}
\caption{The final $\left|d_2^T\right|^2$ projection for the static $F_z$ offset (the slowly-variable noise regime) -- solid curve. The results for the maximal $f_c$ considered in the RTN model are shown for comparison. The points are simulation data and the dotted line is a $1/(1 + c F_N^2)$ function fitted to the data.\label{static_shift}}
\end{figure}

The results for the static offset are presented in Fig.~\ref{static_shift}. The value of the final projection was averaged over both possible signs of $F_N$. The effectiveness of the operation is nearly one for $F_N=0$ and it drops to nearly-zero as $F_N$ increases. The condition for high fidelity $\left|d_2^T(T_P)\right|^2>0.9$ is in approximation $F_N < 0.05$~kV/m and if $F_N > 0.13$~kV/m, then the probability of a successful operation is less than a half. These relations may be seen as estimates for the necessary conditions of an electric field control in any experimental realization of the presented scheme.

The imperfections in the $F_z$ control process can be intuitively divided into two categories. The first one can be thought of, in a simplified way, as a systematic error type: instead of the desired $F_1$ value, the system is tuned to $F_1 + \Delta$ at a given time and the \textit{lifetime} of the $\Delta$ error value is large. The second type consists of errors constantly oscillating around the correct $F_1$ value (\textit{i.e.} RTN model). We have shown that the first type of error, the static one, is more destructive to the described process than the second one. Thus the experimental setup used for realization of the proposed scheme should especially minimize the systematic kind of $F_z$ control error.

\section{Discussion}
The values of the parameters of any given realization of the quantum dot system are not perfectly known \textit{a priori}. This relates to the exact size and geometry of the dots, the confining potential, the precise value of the $g$-factor, among others. Fortunately, the presented scheme is -- on the general level -- adaptive to the specifics of a given system. The only necessary condition is that the four lowest-lying eigenstates of a system need to be qualitatively similar to the ones presented in Fig. \ref{Hs_spectrum}(b) for some magnetic field $B_z$ and some electric field $F_z$ range. That is, the two avoided crossings used for the transitions need to be present in the energy spectrum. Unfortunately, the efficiency of the whole process is strongly dependent on the specifics of the driving impulse, and these specifics depend in turn on the details of the previously mentioned parameters of the system. Thus, for a practical realization of this idea, it is necessary to study a given system experimentally in order to establish reliable estimates of the parameters. This is especially true in the case of the characteristic energy of the avoided crossing C \mbox{in Fig. \ref{Hs_spectrum}(b).} Any imperfections in the axial symmetry of the nanowire shape as well as the piezoelectric effects (also breaking this symmetry) will contribute to the mixing of the states with different $J_z$ values. In practice, it would be most efficient to take the approach of Ref. \onlinecite{LZSM_elec_theory}, i.e. to treat the anticrossing energy as a fittable parameter and try to deduce its value from experimental data. After that, one can employ the presented $\lbrace\alpha,\beta,\gamma,\delta,\epsilon\rbrace$ scheme and optimize each of the steps and join them together as has been presented above.

\section{Summary and Conclusions}
In the presented work a system of double quantum dots, created in an InSb nanowire by application of an external potential, was investigated. The energy spectrum of the system in a static electric field, applied in the direction of the wire, was obtained. The presence of the non-axially symmetric terms in the overall Hamiltonian leads to the formation of an avoided crossing in the spectrum, which involves two states of opposite spin states, in addition to the tunnel-coupling one. A scheme for reversing the spin state of a hole by manipulating the evolution with electric field was proposed, based on driving the hole state through two anticrossings. The results provided show that a perfect realization of the process, with an exact control over the electric field, is possible and the total process time is one order of magnitude smaller than the realization time of an alternative \textit{classic} EDSR approach. The impact of an imperfect control of the driving factor was studied with two simple models that correspond to two different kinds of error, respectively.

\appendix
\section{ The transformation of Kohn-Luttinger and Dresselhaus Hamiltonians from $\langle100\rangle$ to $\langle111\rangle$ direction\label{AppA}}
The 4-band Kohn-Luttinger Hamiltonian for the $\langle100\rangle$ crystal orientation is given by \mbox{Eq. (\ref{KL100})} and the Dresselhaus spin-orbit Hamiltonian in this orientation is given by \mbox{Eq. (\ref{Dr_100})}. Please note, that both are defined in the terms of $\vec{k}$ and $\vec{J}$ vectors. To transform the Hamiltonians to the $\langle111\rangle$ crystal orientation, one needs to express the coordinates of these vectors for $\langle100\rangle$ in terms of their coordinates for the $\langle111\rangle$ orientation. Within the scope of this appendix, the $x, y, z, k_x, k_y, k_z, J_x, J_y, J_z$ symbols are used for the \textit{old} crystal orientation and the $x^{'}, y^{'}, z^{'}, k_x^{'}, k_y^{'}, k_z^{'}, J_x^{'}, J_y^{'}, J_z^{'}$ symbols are used for the \textit{new} one.

In $x, y, z$ coordinates, the $x^{'}$ axis goes along the $[1,1,-2]$ vector, the $y^{'}$ along the $[-1,1,0]$ vector and $z^{'}$ along the $[1,1,1]$ vector. The related versors are:
\begin{equation}
\begin{cases}
\hat{x}^{'} = \frac{\hat{x}}{\sqrt{6}} + \frac{\hat{y}}{\sqrt{6}} - \frac{2 \hat{z}}{\sqrt{6}} \\
\hat{y}^{'} = -\frac{\hat{x}}{\sqrt{2}} + \frac{\hat{y}}{\sqrt{2}} \\
\hat{z}^{'} = \frac{\hat{x}}{\sqrt{3}} + \frac{\hat{y}}{\sqrt{3}} + \frac{\hat{z}}{\sqrt{3}} 
\label{xinxp}
\end{cases}
\end{equation}
and in consequence:
\begin{equation}
\begin{cases}
x = \frac{x^{'}}{\sqrt{6}} - \frac{y^{'}}{\sqrt{2}} + \frac{z^{'}}{\sqrt{3}} \\
y = \frac{x^{'}}{\sqrt{6}} + \frac{y^{'}}{\sqrt{2}} + \frac{z^{'}}{\sqrt{3}} \\
z = -\sqrt{\frac{2}{3}} x^{'} + \frac{z^{'}}{\sqrt{3}} 
\end{cases}.
\end{equation}

The $u$-derivative in the $\lbrace x^{'}, y^{'}, z^{'} \rbrace$ basis is equal to \mbox{$\frac{\partial}{{\partial}u} = \frac{{\partial}x^{'}}{{\partial}u} \frac{\partial}{{\partial}x^{'}} + \frac{{\partial}y^{'}}{{\partial}u} \frac{\partial}{{\partial}y^{'}} + \frac{{\partial}z^{'}}{{\partial}u} \frac{\partial}{{\partial}z^{'}}$.} After taking into account \mbox{Eq. (\ref{xinxp}),} the following relation for $\vec{k}$ is obtained:
\begin{equation}
\begin{cases}
k_x = \frac{1}{\sqrt{6}} k_x^{'} - \frac{1}{\sqrt{2}} k_y^{'} + \frac{1}{\sqrt{3}} k_z^{'}\\
k_y = \frac{1}{\sqrt{6}} k_x^{'} + \frac{1}{\sqrt{2}} k_y^{'} + \frac{1}{\sqrt{3}} k_z^{'}\\
k_z = -\sqrt{\frac{2}{3}} k_x^{'} + \frac{1}{\sqrt{3}} k_z^{'}
\end{cases}.
\end{equation}

The spin vector in the old basis is equal to:
\begin{equation}
\vec{\sigma} = \left[J_x, J_y, J_z\right] = J_x \hat{x} + J_y \hat{y} + J_z \hat{z}
\label{sigma1}
\end{equation}
and in the new basis the same vector is given by:
\begin{equation}
\vec{\sigma} = \left[J_x^{'}, J_y^{'}, J_z^{'}\right] = J_x^{'} \hat{x}^{'} + J_y^{'} \hat{y}^{'} + J_z^{'} \hat{z}^{'}.
\label{sigma2}
\end{equation}
By comparing the right hand sides of \mbox{Eqs. (\ref{sigma1})} and (\ref{sigma2}) and taking into account \mbox{Eq. (\ref{xinxp})}, the following expression for spin matrices can be obtained:
\begin{equation}
\begin{cases}
J_x = \frac{1}{\sqrt{6}} J_x^{'} - \frac{1}{\sqrt{2}} J_y^{'} + \frac{1}{\sqrt{3}} J_z^{'}\\
J_y = \frac{1}{\sqrt{6}} J_x^{'} + \frac{1}{\sqrt{2}} J_y^{'} + \frac{1}{\sqrt{3}} J_z^{'}\\
J_z = -\sqrt{\frac{2}{3}} J_x^{'} + \frac{1}{\sqrt{3}} J_z^{'}
\end{cases}.
\end{equation}

\section{ The Dresselhaus Hamiltonian matrices for specific $J_z$ values\label{AppB}}
The Dresselhaus Hamiltonian for the considered system [see Eqs. (\ref{Dr_111}), (\ref{Dr_111_oper}) and Table \ref{Dress_constants}] is defined in terms of $\hat{k}_{-},\hat{k}_{+},\hat{k}^{2}_{\perp}$ and $\hat{k}_{z}$ operators that act on specific types of valence bands. Each single-band component of a $\hat{H}_s$ eigenvector has a $e^{i J^{en}_{z} \phi}$ type of angular dependency and thus a defined envelope angular momentum quantum number $J^{en}_{z}$ [see Eqs. (\ref{ho1peigen}) and (\ref{oneparticle_symm})]. In the case of matrix elements of the mentioned operators, for states with $(J^{en}_{z})_L$ and $(J^{en}_{z})_R$ quantum numbers, for ${\mid}L{\rangle}$ and ${\mid}R{\rangle}$ states respectively, the result is nonzero only in the case of some relations of these numbers:
\begin{eqnarray}
(J^{en}_{z})_R~{\neq}~(J^{en}_{z})_L+1 &{\implies}& {\langle}L{\mid}\hat{k}_{-}{\mid}R{\rangle} = 0, \nonumber\\
(J^{en}_{z})_R~{\neq}~(J^{en}_{z})_L-1 &{\implies}& {\langle}L{\mid}\hat{k}_{+}{\mid}R{\rangle} = 0, \nonumber\\
(J^{en}_{z})_R~{\neq}~(J^{en}_{z})_L &{\implies}& {\langle}L{\mid}\hat{k}^{2}_{\perp}{\mid}R{\rangle} = 0, \nonumber\\
(J^{en}_{z})_R~{\neq}~(J^{en}_{z})_L &{\implies}& {\langle}L{\mid}\hat{k}_{z}{\mid}R{\rangle} = 0.
\end{eqnarray}
The mentioned relations lead to a significant simplification of the Dresselhaus Hamiltonian for each specific pair of~$J_z$ numbers.

If we take into account only the states with a certain~$J_z$, then the four envelope angular momentum quantum numbers~$J^{en}_{z}$ in \mbox{$(HH\uparrow, LH\downarrow, LH\uparrow, HH\downarrow)$} basis are $(J_z-\frac{3}{2},J_z+\frac{1}{2},J_z-\frac{1}{2},J_z+\frac{3}{2})$. For example, the $J_z=-\frac{3}{2}$ gives $(-3,-1,-2,0)$ and for $J_z=+\frac{3}{2}$ the values $(0,2,1,3)$ are obtained. The effective form of the Hamiltonian, for equal $J_z$ numbers of \textit{bra} and \textit{ket} states, becomes:
\begin{eqnarray}
\hat{H}_{D}^{'} &=& \left(
\begin{array}{cccc}
0 & \hat{O}_{3} & \hat{O}_{2} & \hat{O}_{4} \\
\hat{O}_{3}^{+} & 0 & \hat{O}_{5} & \hat{O}_{2} \\ 
\hat{O}_{2}^{+} & \hat{O}_{5}^{+} & 0 & -\hat{O}_{3} \\
\hat{O}_{4}^{+} & \hat{O}_{2}^{+} & -\hat{O}_{3}^{+} & 0
\end{array}\right),
\label{Dr_same}
\end{eqnarray}
with the operators defined as follows:  
\begin{eqnarray}
\hat{O}_{2} &=& -\frac{C_k}{\sqrt{3}} (i \hat{k}_{-}) + c_{2} \hat{k}^{2}_{\perp} (i \hat{k}_{-}) + c_{4} \hat{k}_z^2 (i \hat{k}_{-}),\nonumber\\
\hat{O}_{3} &=& - i \sqrt{3} c_{6} \hat{k}_z (i \hat{k}_{-})^{2},\nonumber\\
\hat{O}_{4} &=& c_{6} (i \hat{k}_{-})^3, \nonumber\\
\hat{O}_{5} &=& -C_k (i \hat{k}_{+}) - c_{10} \hat{k}^{2}_{\perp} (i \hat{k}_{+}) - c_{12} \hat{k}_z^2 (i \hat{k}_{+}),\nonumber\\
\hat{O}_{2}^{+} &=& \frac{C_k}{\sqrt{3}} (i \hat{k}_{+}) - c_2 \hat{k}^{2}_{\perp} (i \hat{k}_{+}) - c_4 \hat{k}_z^2 (i \hat{k}_{+}),\nonumber\\
\hat{O}_{3}^{+} &=& i \sqrt{3} c_{6} \hat{k}_z (i \hat{k}_{+})^2,\nonumber\\
\hat{O}_{4}^{+} &=& -c_{6} (i \hat{k}_{+})^3 , \nonumber\\
\hat{O}_{5}^{+} &=& C_k (i \hat{k}_{-}) + c_{10} \hat{k}^{2}_{\perp} (i \hat{k}_{-}) + c_{12} \hat{k}_z^2 (i \hat{k}_{-}),
\label{Dr_same_oper}
\end{eqnarray}
and the constants $c_2$ to $c_{12}$ as defined in Table \ref{Dress_constants}.

Analogously, the effective Hamiltonian for $J_z$ and \mbox{$J_z+3$} (for \textit{bra} and \textit{ket} states, respectively) can be obtained. This corresponds, for example, to a pair of $J_z=-\frac{3}{2}$ and $J_z=+\frac{3}{2}$ states with the sets of envelope angular momentum quantum numbers~$J^{en}_{z}$ of $(-3,-1,-2,0)$ and of $(0,2,1,3)$, respectively. In this case, the effective Hamiltonian takes the form of:
\begin{eqnarray}
\hat{H}_{D}^{''} &=& \left(
\begin{array}{cccc}
\hat{O}_{1} & 0 & 0 & 0 \\
\hat{O}_{3}^{+} & -a\hat{O}_{1} & \hat{O}_{5} & 0 \\ 
\hat{O}_{2}^{+} & 0 & a\hat{O}_{1} & 0 \\
\hat{O}_{4}^{+} & \hat{O}_{2}^{+} & -\hat{O}_{3}^{+} & -\hat{O}_{1}
\end{array}\right),
\label{Dr_+3}
\end{eqnarray}
with the operators defined as follows:  
\begin{eqnarray}
\hat{O}_{1} &=& -c_{1} (i \hat{k}_{-})^3,\nonumber\\
\hat{O}_{5} &=& i c_{11} \hat{k}_z (i \hat{k}_{-})^2,\nonumber\\
\hat{O}_{2}^{+} &=& i c_3 \hat{k}_z \hat{K}_1^2,\nonumber\\
\hat{O}_{3}^{+} &=& -\frac{C_k}{\sqrt{6}} \hat{K}_1 - c_5 \hat{K}_3 \hat{K}_1 - c_7 \hat{K}_1 \hat{k}_z^2,\nonumber\\
\hat{O}_{4}^{+} &=& i \sqrt{2} C_k \hat{k}_z + i c_{13} \hat{k}_z \hat{K}_3 + i c_9 \hat{k}_z^3,
\label{Dr_+3_oper}
\end{eqnarray}
and the constants $a$ and $c_1$ to $c_{13}$ as defined in Table \ref{Dress_constants}.

Please note that in the cases of I) $J_z$ and \mbox{$J_z^{'}=J_z+1$} (\textit{e.g} the \mbox{$J_z=-\frac{5}{2}$}, \mbox{$J_z^{'}=-\frac{3}{2}$} pair and the \mbox{$J_z=\frac{1}{2}$}, \mbox{$J_z^{'}=\frac{3}{2}$} pair), II) $J_z$ and \mbox{$J_z^{'}=J_z+2$} (\textit{e.g} the \mbox{$J_z=-\frac{3}{2}$}, \mbox{$J_z^{'}=\frac{1}{2}$} pair), and III) $J_z$ and \mbox{$J_z^{'}=J_z+4$} (\textit{e.g} the \mbox{$J_z=-\frac{4}{2}$}, \mbox{$J_z^{'}=\frac{3}{2}$} pair), the effective Dresselhaus Hamiltonian is zero and hence Dresselhaus-type spin-orbit does not induce mixing of these states.

\section{ Detailed optimization of the driving signal\label{AppC}}

\subsubsection{Transfer using the larger avoided crossing}

For this part of the transfer an initial state identical to the ground state $d_1^T(0)=1$ is assumed and our goal is to maximize the first excited state projection after the evolution $\left|d_2^T(T_P)\right|^2$ [see Eq. (\ref{time_evolution_basis})]. The following function was accepted as the driving element for the $\beta$ transition [see Fig.~\ref{Hs_spectrum}(b)]:
\begin{equation}
F_z(t) = \frac{p_s + p_b}{2} + \frac{p_s-p_b}{2}\,\cos\left(\frac{6 \pi t}{T_P}\right),
\label{F_z(t)_large}
\end{equation}
where $p_s$ is the initial and final value of the $F_z$ pulse, $p_b$ is the bouncing point on the $F_z<0$ side of the avoided crossing and $T_P$ is the evolution time. The function Eq.~(\ref{F_z(t)_large}) has three parameters: $p_s$, $p_b$ and $T_P$. A three-dimensional optimization of this parameters has been done in order to maximize the efficiency of the transfer. The method used was grid search with a mesh spacing of ${\Delta}p_s={\Delta}p_b=0.01$ kV/m and with ${\Delta}{\text{Log}}_{10}(\frac{T_P}{\text{1 s}})=0.01$. The ranges for the search were chosen so that the values of $p_s$ and $p_b$ ensure the correct overlap of the $F_z(t)$ pulse range and the range of the A anticrossing. The evolution time $T_P$ corresponds to the timescale of the process. The points on the search grid were chosen to lie in equal distances of ${\text{Log}}_{10}(\frac{T_P}{\text{1 s}})$ because of the need to search among more than one order of magnitude. The obtained values of the parameters are: \mbox{$p_s=1.16$ kV/m}, \mbox{$p_b=-1.98$ kV/m} and \mbox{$T_P=295$ ps}. The optimal value $|d_2^T(T_P)|^2>0.996$ was obtained. This value is very close to unity and it is sufficient for the realization of the intended goal. The range of the pulse is marked in Fig.~\ref{Hs_spectrum}(b) as $F_A$.

\begin{figure}[!ht]
\rotatebox{0}{\epsfxsize=70mm \epsfbox[80 35 795 535]{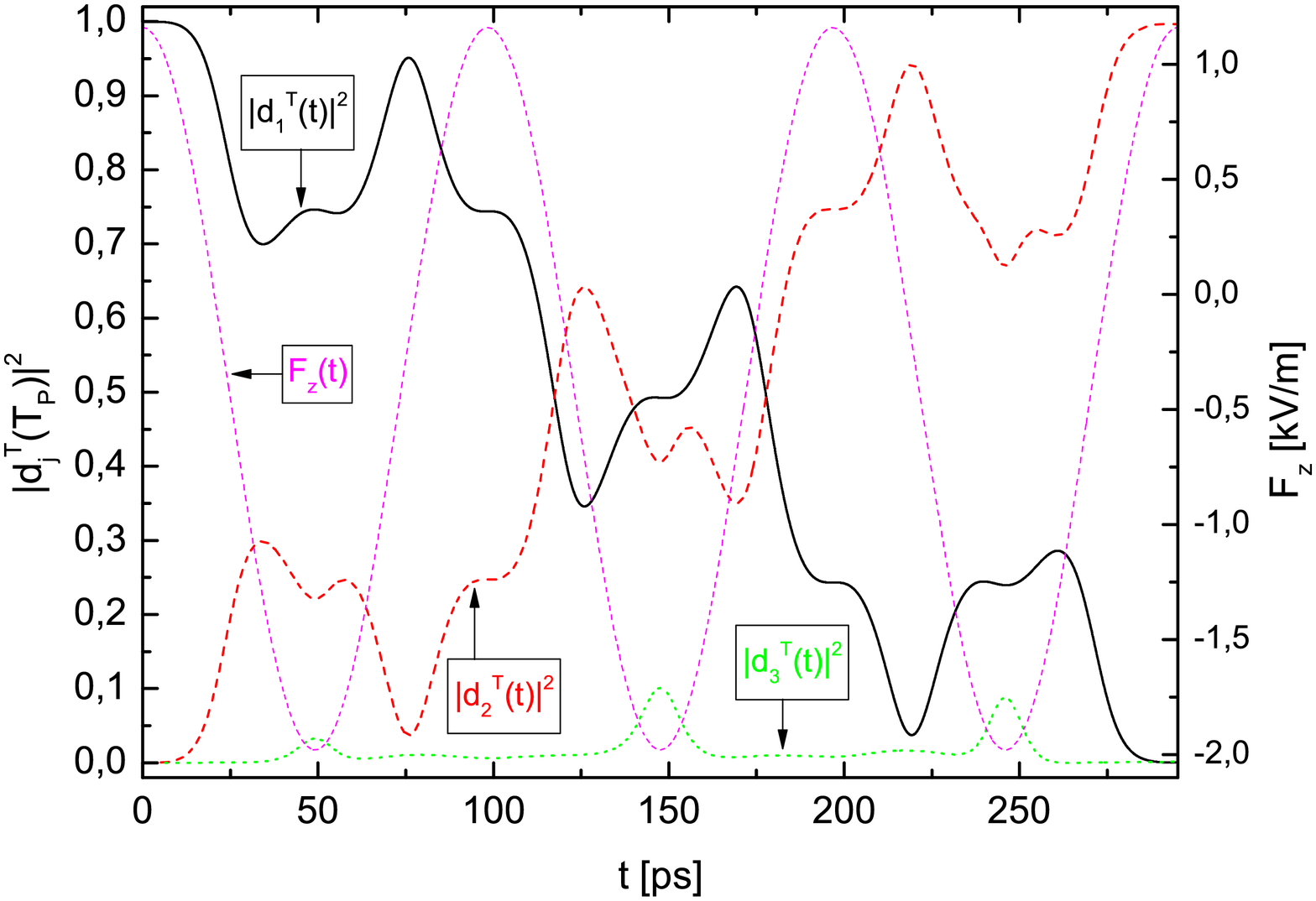}}
\caption{Evolution for $\beta$ transition; \mbox{$p_s=1.16$ kV/m}, \mbox{$p_b=-1.98$ kV/m} and $T_P=295$ ps. Left axis: the $|d_j^T(t)|^2$ projections for the time-dependent state $\Phi(\vec{r},t)$ on the time-independent states \mbox{$\Psi_{j}(\vec{r})$}. For $j=4$ the projection is not shown as it is negligible for all $t{\in}\left(0,T_P\right)$. Right axis: the $F_z(t)$ electric field drive.\label{evolution_beta}}\end{figure}

The evolution for the set of parameters obtained in the optimization is presented in Fig. \ref{evolution_beta}. The evolving state starts in the ground state: $d_1^T(0)=1$. As the impulse $F_z(t)$ starts to diverge from the initial value $p_s$, the $|d_1^T|^2$ drops and $|d_2^T|^2$ grows by an equal amount. In the regions, where $F_z$ is close to $p_b$, the second excited state is active in the process with $|d_3^T(0)|^2>0$ \textit{storing} a bit of the evolving wavefunction for a small moment. This is due to the fact that $p_b$ lies relatively close to the C anticrossing that involves the first and the second excited levels [see $F_A$ in Fig.~\ref{Hs_spectrum}(b)]. As the impulse begins to go back from $p_b$ to $p_s$, the transfer continues. The process repeats three times until the evolving state is transferred almost completely to the first excited time-independent level: $|d_2^T(T_P)|^2{\approx}1$. The first movement through the avoided crossing from $p_s$ to $p_b$ and back results in transferring $0.25$ of the initial $|d_1^T|^2$ projection to the $|d_2^T|^2$ one. The second go increases $|d_2^T|^2$ by additional $0.5$ and the last one results in transferring the remaining $0.25$ from $|d_1^T|^2$.

\subsubsection{Transfer using the smaller avoided crossing}

At this stage the initial state of the system is equal to the first excited time-independent state: $d^T_2(0) = 1$. We seek to maximize $\left|d^T_2(T_P)\right|^2$, that is the projection of the same kind after the evolution [see Eq. (\ref{time_evolution_basis})]. The time-dependent electric field for the $\delta$ transition, see Fig.~\ref{Hs_spectrum}(b), is given by:
\begin{equation}
F_z(t) = \frac{p_s + p_f}{2} + \frac{p_s-p_f}{2}\,\cos\left(\frac{5 \pi t}{T_P}\right),
\label{F_z(t)_small}
\end{equation}
where $p_s$ is the initial value of the $F_z$ pulse, $p_f$ is the final value and $T_P$ is the evolution time. The function Eq.~(\ref{F_z(t)_small}) has three parameters: $p_s$, $p_f$ and $T_P$. A three-dimensional optimization of this parameters has been done in order to maximize the efficiency of the transfer, with the grid search method similar to the one described above, for the case of the larger anticrossing. The three-dimensional optimization yielded the following values for the parameters: \mbox{$p_s=1.23$ kV/m}, \mbox{$p_f=2.56$ kV/m} and \mbox{$T_P=240$ ps}. The optimal value $|d_2^T(T_P)|^2>0.995$ was obtained. This value is very close to unity and it is sufficient for the realization of the intended goal. The range of the pulse is shown in Fig.~\ref{Hs_spectrum}(b) as the $\delta$ stage.

\begin{figure}[!ht]
\rotatebox{0}{\epsfxsize=70mm \epsfbox[80 35 795 535]{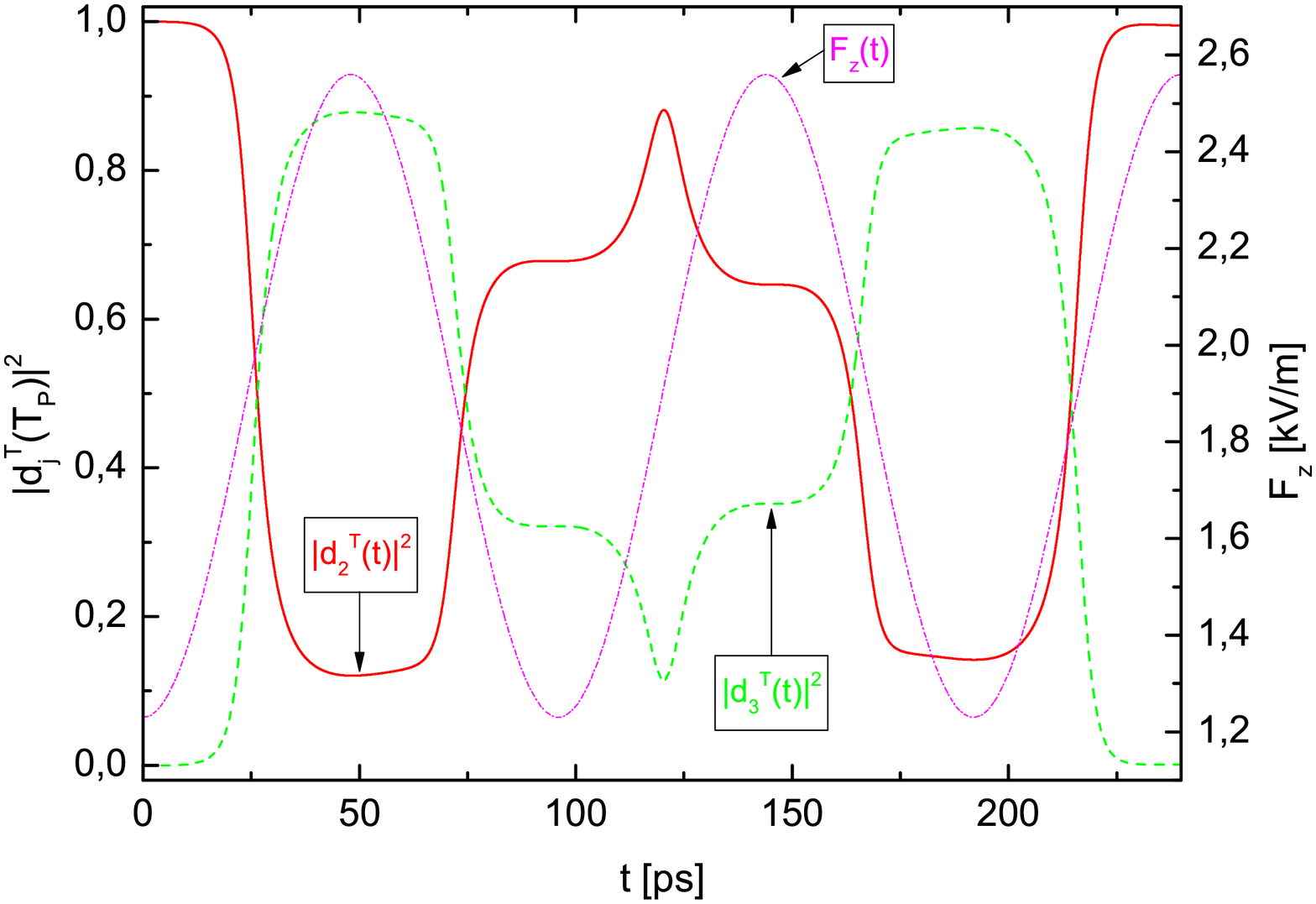}}
\caption{Evolution for $\delta$ transition; \mbox{$p_s=1.23$ kV/m}, \mbox{$p_f=2.56$ kV/m} and $T_P=240$ ps. Left axis: the $|d_j^T(t)|^2$ projections for the time-dependent state $\Phi(\vec{r},t)$ on the time-independent states \mbox{$\Psi_{j}(\vec{r})$}. For $j{\in}\lbrace1,4\rbrace$ the projections are not shown as they are negligible for all $t{\in}\left(0,T_P\right)$. Right axis: the $F_z(t)$ electric field drive.\label{evolution_delta}}\end{figure}

The evolution for the set of parameters obtained from the optimization is presented in Fig. \ref{evolution_delta}. The evolving state starts in the first excited state $d_2^T(0)=1$. The driving field $F_z(t)$ has a minimal value $p_s = 1.23$ kV/m for \mbox{$t\in\lbrace0,$~$96$~$\text{ps},$~$192$~$\text{ps}\rbrace$}, which corresponds to the hole being driven to the \textit{left} side of the avoided crossing D in Fig. \ref{Hs_spectrum}(b). For this sequence of time values a characteristic behaviour can be observed: the corresponding values of $|d_2^T|^2$ are systematically decreasing and the corresponding values of $|d_3^T|^2$ are systematically increasing. On the other hand, when the driving field $F_z(t)$ has maximal value $p_f = 2.56$ kV/m for $t\in\lbrace48$~$\text{ps},$~$144$~$\text{ps},$~$T_P\rbrace$, the hole system is on the \textit{right} side of the mentioned avoided crossing. For this sequence of time values the corresponding values of $|d_2^T|^2$ are systematically increasing and the corresponding values of $|d_3^T|^2$ are systematically decreasing. This marks the transition from the system in which hole occupies the first excited state \mbox{$\Psi_{2}(\vec{r})$} on the \textit{left} side and it occupies the second excited state \mbox{$\Psi_{3}(\vec{r})$} on the \textit{right} side, to the system with the reversed occupation characteristics.

\subsubsection{$\alpha$, $\gamma$ and $\epsilon$ transfers}

The optimization of the transfers through the both avoided crossings yields the border $F_z$ values for other stages of the process, see Fig. \ref{Hs_spectrum}(b). Explicitly, the $\alpha$ transfer should drive the system from $F_z=4$ kV/m to $F_z=1.16$ kV/m, the $\gamma$ transfer needs to initiate at $F_z=1.16$ kV/m and end at $F_z=1.23$ kV/m, and the $\epsilon$ transfer is to be done between electric field values of $F_z=2.56$ kV/m and $F_z=4$ kV/m. For each of these stages, the hole state occupation should remain the same. The only limitation of the process is that it should be done slowly enough to enable the wavefunction to accommodate for the change in $F_z$. In other words, these transfers should be made quasi-adiabatically.

\begin{figure}[!ht]
\epsfxsize=70mm \epsfbox[50 35 725 525] {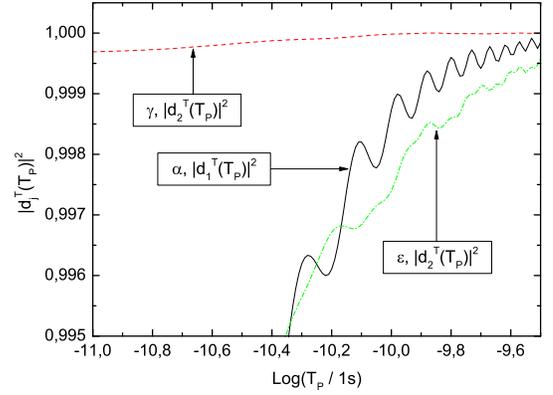}
\caption{Optimization of the $T_P$ parameter for $\alpha$, $\gamma$ and $\epsilon$ transitions. The~$\left|d_j^T(T_P)\right|^2$ projection for the time-dependent state $\Phi(\vec{r},t)$ on the $j$-th time-independent state \mbox{$\Psi_{j}(\vec{r})$} as a function of evolution time $T_P$, where $j=1$ for $\alpha$ and $j=2$ for $\gamma$ and $\epsilon$. \label{transfer_TP_other}}
\end{figure}

The following $F_z(t)$ function driving the system from $p_s$ to $p_f$ has been adopted:
\begin{equation}
F_z(t) = p_s + (p_f-p_s) \frac{t}{T_P},
\label{F_z(t)_other}
\end{equation}
which (if the $p_s$ and $p_f$ values are already set) has only one free parameter $T_P$, i.e. the evolution time. The results of optimization of this parameter for $\alpha$, $\gamma$ and $\epsilon$ transfers are presented in Fig. \ref{transfer_TP_other}. The $T_P$ should be as small as possible to make the process fast, while also should guarantee very good transfer effectiveness. The evolution time adopted for the $\alpha$ step is $T_P = 10^{-9.9} \text{~s} = 125 \text{~ps}$, for $\gamma$ step it is $T_P = 10^{-10.3} \text{~s} = 50 \text{~ps}$ and for $\epsilon$ step it is $T_P = 10^{-9.825} \text{~s} = 150 \text{~ps}$. In each case the chosen time allows for the transition effectiveness $> 0.995$.

\section{ The asymmetric system\label{AppD}}

\begin{figure*}[!ht]
\begin{tabular}{ccc}
(a) & (b) & (c)\\
\rotatebox{0}{\epsfxsize=55mm \epsfbox[55 40 680 540] {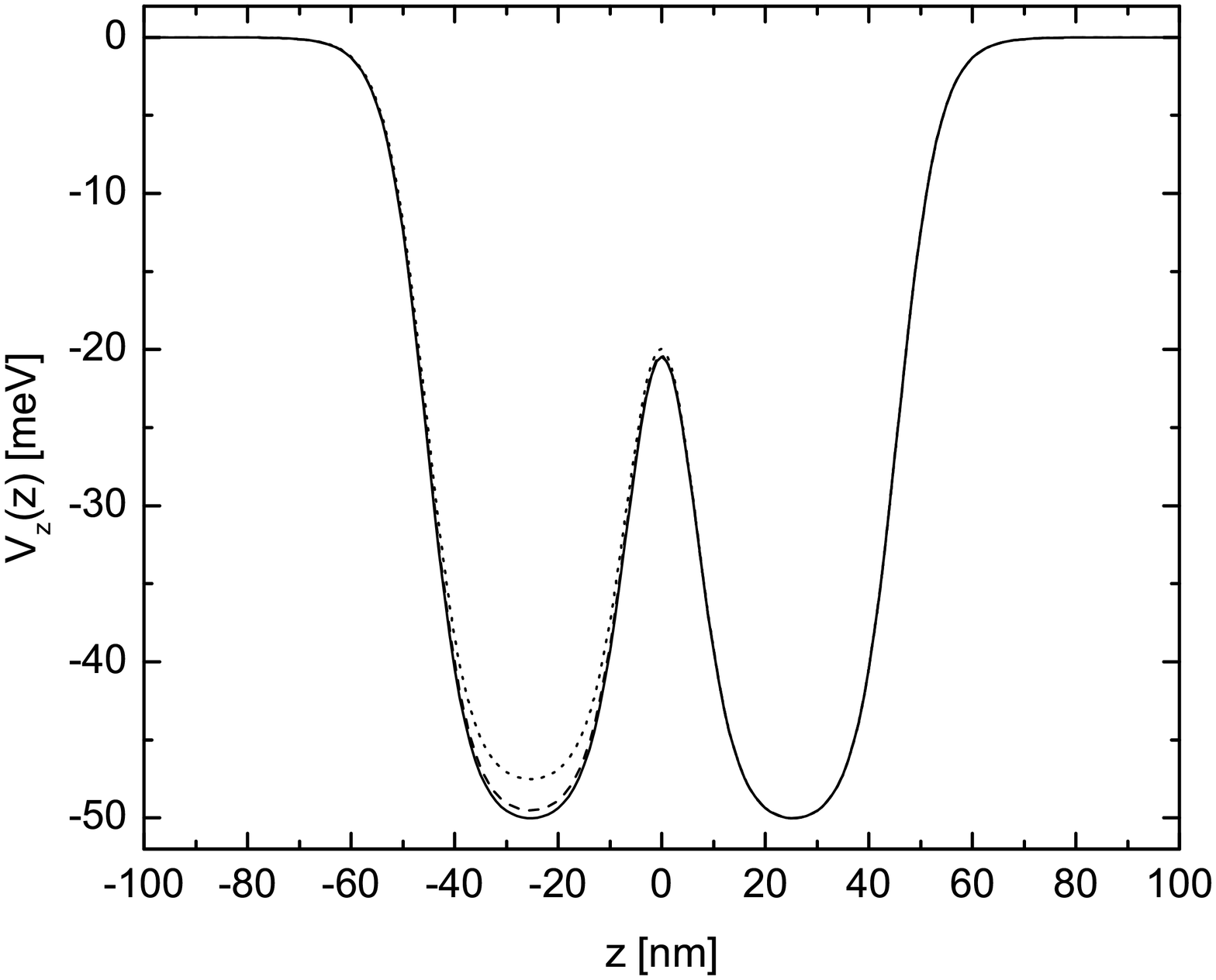}} & 
\rotatebox{0}{\epsfxsize=55mm \epsfbox[55 40 680 540] {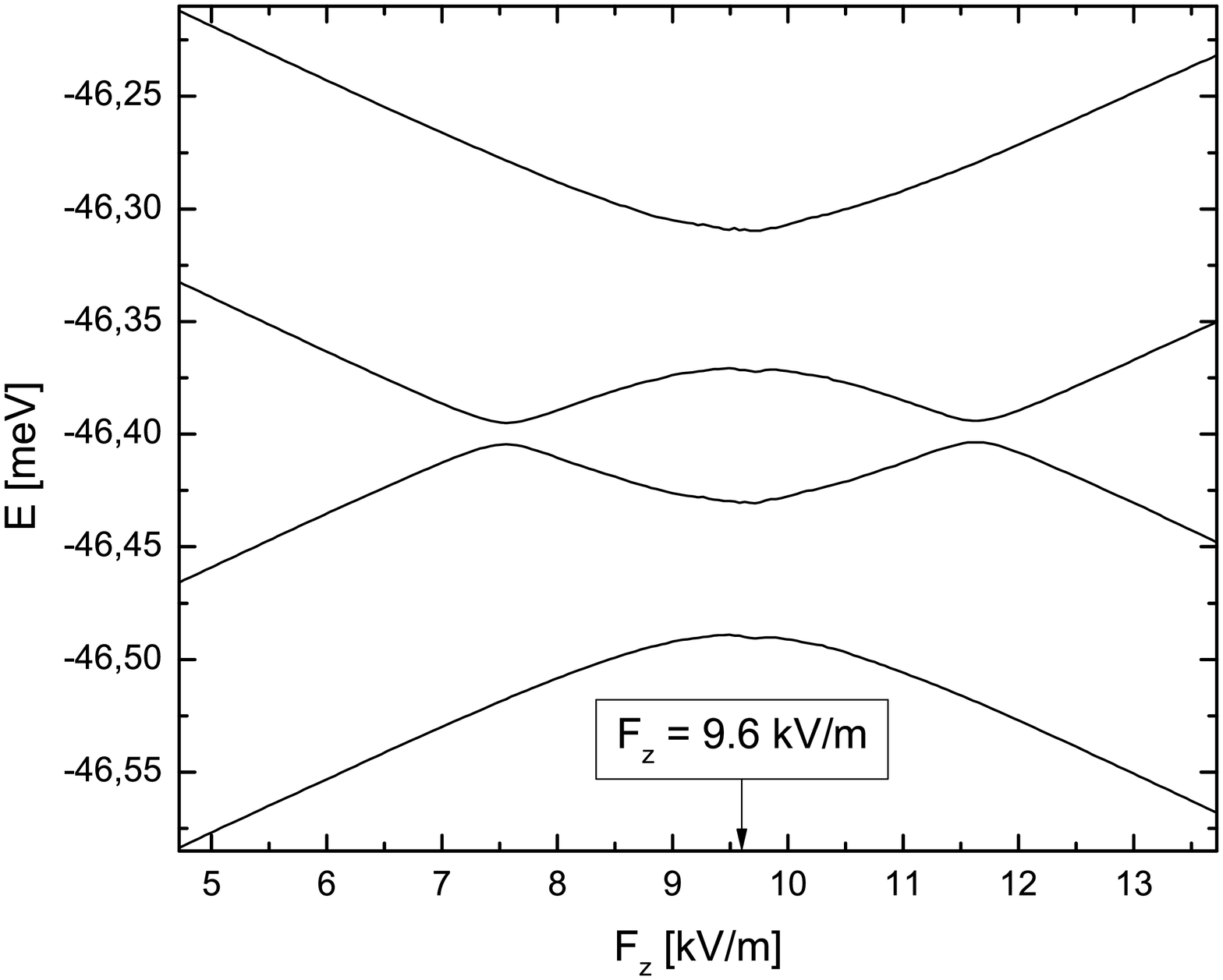}} &
\rotatebox{0}{\epsfxsize=55mm \epsfbox[55 40 680 540] {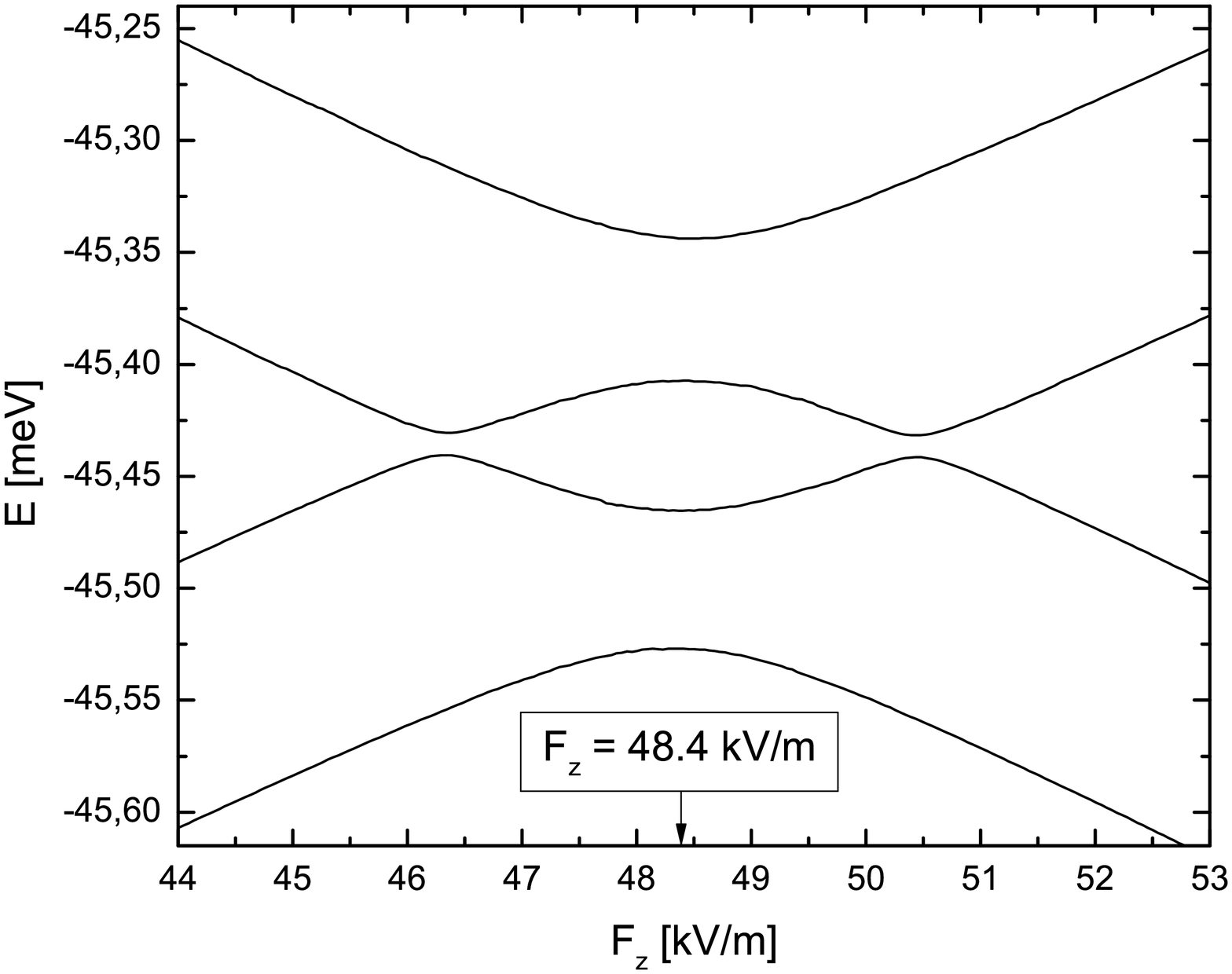}}
\end{tabular}
\caption{(a) The confinement potential along the $z$ axis with: no asymmetry -- solid line, $1\%$~asymmetry -- dashed line, $5\%$~asymmetry -- dotted line. (b),(c) Hole energy spectrum of the system with non-axially-symmetric terms included with: \mbox{(b) $1\%$~asymmetry,} \mbox{(c) $5\%$ asymmetry.}\label{asymmetric}}
\end{figure*}

The quantum dot confinement, that was considered in the main part of the work, is symmetric in respect to reversing the $z$ axis. In reality, it is probable that the perfect symmetry in this regard cannot be maintained, due to technological limitations. In this appendix, we present the study of an impact of a small asymmetry in the confinement potential on the behavior of the system. Two cases have been considered: I) the $z<0$ dot is $1\%$ shallower than the $z>0$ dot, and II) the $z<0$ dot is $5\%$ shallower than the $z>0$ dot. The $z$ axis confining potential has the form of:
\begin{equation}
V_z(z) = V_0 (x_{as} V^{d}_z(z_{l}) + V^{d}_z(z_{r})),
\end{equation}
where $x_{as}\in\lbrace0.95, 0.99\rbrace$ and the other symbols are as defined in Eq. (\ref{confinement}).

The shapes of the $z$-axis confinement potentials are presented in Fig. \ref{asymmetric}(a) and the energy spectra of the system, with the non-axial part taken into consideration, are presented in Fig. \ref{asymmetric}(b) and (c). The potential with the smaller asymmetry is almost identical to the symmetric one and the potential with the bigger asymmetry is easily discernible from the symmetric one, see Fig. \ref{asymmetric}(a). However, the spectra in both cases have nearly the same character as the one for the symmetric system, compare Fig. \ref{Hs_spectrum}(b) and Fig. \ref{asymmetric}(b) and (c). The only difference is the shift of the whole spectrum in the terms of the electric field. The position of the middle of the tunneling anticrossing is at about $F_z=0$ for the system considered in the main part of the work, but it is shifted to about $F_z=9.6$ kV/m in the case of the $1\%$ asymmetry [see Fig. \ref{asymmetric}(b)] and to $F_z=48.4$ kV/m in the case of the $5\%$ asymmetry [see Fig. \ref{asymmetric}(c)]. The results the for case when the $z>0$ dot is shallower than the $z<0$ one (not shown) are nearly the same, with the exception that the electric field shift is negative. In conclusion, the impact of the small asymmetry on the behavior of the system is minimal and the evolution research can be done for the symmetric system.

\section{ The analogy between the high-frequency RTN regime and the motional narrowing\label{AppE}}

\begin{figure}[!ht]
\epsfxsize=70mm \epsfbox[60 35 680 540]{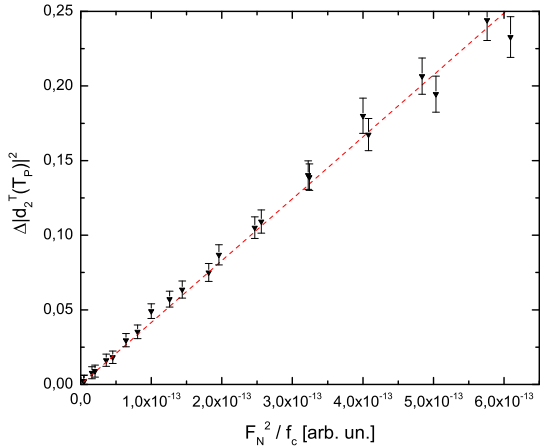}
\caption{The loss in final $\left|d_2^T\right|^2$ projection for the RTN noise simulation as a function of the $\frac{F_N^2}{f_c}$ argument. The line is a linear fit to the data.\label{theory-high}}
\end{figure}

The motional narrowing is a decrease in the linewidth of a resonant frequency, that is an effect of the motion in an inhomogeneous system. For the description of the phenomenon in the case of magnetic resonance, see e.g. Ref. \onlinecite{motional}. In a very simple model it can be described by the following formula (Ref. \onlinecite{motional}, page 213):
\begin{equation}
\frac{1}{T_2} = \gamma_n^2 H_z^2 \tau
\end{equation}
where: $H_z$ is the magnetic field amplitude, $\tau$ is the lifetime of a $\vec{H}$ orientation, $T_2$ is the relaxation time, and $\gamma_n^2$ is a constant. In our system, the analog of $H_z$ is the RTN amplitude of the electric field $F_N$. The equivalent of reverse of $\tau$ is the noise frequency $f_c$. As a first approximation, we assume that \textit{lhs} $\frac{1}{T_2}$ corresponds to the final efficiency loss of the transfer $\Delta\left|d_2^T(T_P)\right|^2$. This leads to the following equation:
\begin{equation}
\Delta\left|d_2^T(T_P)\right|^2 = C \frac{F_N^2}{f_c}.
\end{equation}
The results of the RTN simulation for the highest frequency and small amplitude regime, presented as a function of the $\frac{F_N^2}{f_c}$ argument, are given in Fig. \ref{theory-high}. The corresponding area is marked in Fig. \ref{noise} with a square. The linear function fits the data quite well for the considered range of parameters.


\begin{thebibliography}{00}
\bibitem{annotation1} The explicit formula for this function is
\begin{eqnarray}
s_p(z) = &-&\frac{818}{z}+57.6-1.60 z+2.34{\cdot}10^{-2} z^2\nonumber\\
&-&1.87{\cdot}10^{-4} z^3+7.72{\cdot}10^{-7} z^4-1.29{\cdot}10^{-9} z^5.\nonumber
\end{eqnarray}
\bibitem{annotation2} The maximal value of $B_z$ considered is $200$ mT, for which the additional confinement potential would be equal to \mbox{$0.11$ meV} for heavy holes and \mbox{$0.04$ meV} in the case of light holes. These values can be omitted when compared with the lateral confinement resulting from the wire-vacuum boundary.
\end{thebibliography}
\end{document}